\documentclass[onecolumn,noshowpacs,superscriptaddress,nobibnotes,nofootinbib,12pt,showkeys,preprintnumbers]{revtex4-1}
\UseRawInputEncoding
\usepackage{float}
\usepackage{amsmath,amssymb}
\usepackage{bm}
\usepackage{slashed}
\usepackage{epsfig}
\usepackage{graphicx}
\usepackage{hyperref}
\usepackage{xcolor}
\usepackage{comment}
\usepackage{soul}

\hypersetup{
    colorlinks=true,
    linkcolor=blue,
    filecolor=magenta,      
    urlcolor=blue,
    citecolor=blue
}

\usepackage{tikz}
\usetikzlibrary{decorations.pathmorphing}
\usetikzlibrary{shapes.geometric}
\usetikzlibrary{calc}
\usetikzlibrary{shapes.misc}
\usetikzlibrary{decorations.markings}
\tikzstyle{gluon}=[decorate, decoration={coil,aspect=0.8, amplitude=1.5pt,  segment length=3pt}]
\tikzstyle{lgluon}=[decorate, decoration={coil,aspect=-0.8, amplitude=1.5pt,  segment length=3pt}]

\begin{document}
\title{Cold nuclear matter effects on azimuthal decorrelation in heavy-ion collisions}
\author{N\'estor Armesto}
\email{nestor.armesto@usc.es}
\author{Florian Cougoulic}
\email{florianthibaultmanuel.cougoulic@usc.es}
\author{Bin Wu}
\email{bin.wu@usc.es}
\affiliation{Instituto Galego de F\'isica de Altas Enerx\'ias IGFAE, Universidade de Santiago de Compostela,
E-15782 Galicia-Spain}
\begin{abstract}
The assumption of factorization lies at the core of calculations of medium effects on observables computable in perturbative Quantum Chromodynamics. In this work we examine this assumption, for which we propose a setup to study hard processes and bulk nuclear matter in heavy-ion collisions on the same footing using the Glauber modelling of heavy nuclei. To exemplify this approach, we calculate the leading-order corrections to azimuthal decorrelation in Drell-Yan and boson-jet processes due to cold nuclear matter effects, not considering radiation. At leading order in both the hard momentum scale and the nuclear size, the impact-parameter dependent cross section is found to factorize for both processes. The factorization formula involves a convolution of the hard cross section with the medium-modified parton distributions, and, for boson-jet production, the medium-modified jet function.
\end{abstract}

\maketitle

\section{Introduction}

Factorization in Quantum Chromodynamics~\cite{Collins:1989gx, Sterman:1995fz, Stewart:2003eft, Becher:2014oda} plays a central role in hadron collider physics. While it has been proved in $l^+l^-$ collisions and Deep Inelastic Scattering (DIS), in hadron-hadron collisions it has only been vetted for processes where the final state consists of colorless particles. Besides, situations are known where factorization fails, see e.g.~\cite{Collins:2007nk,Rogers:2010dm,Mulders:2011zt} and references therein.

Factorization is also widely assumed for hard processes in heavy-ion collisions, but it has been analyzed more rarely that in nucleon-nucleon scattering and most often discussed in the context of DIS and photoproduction, see, e.g.,~\cite{Qiu:2001hj,Cao:2020wlm} and refs. therein. In spite of this, its application lies at the core of studies dedicated to the extraction of properties of the bulk matter or medium produced in such collisions using hard probes like jet or large transverse momentum hadron production. Checks of the (lack of) validity of factorization in collisions involving nuclei require the description of both hard processes and bulk matter in a theoretically well-grounded, unified framework. This central issue remains, despite many efforts, an open question.

Parton saturation, also known as the Color Glass Condensate (CGC), provides a promising description of bulk matter (soft and semihard partons) in the early stages of heavy-ion collisions~\cite{Kovchegov:2012mbw}. In this framework, the production of these partons has been conventionally calculated using classical color sources, as originally proposed in the McLerran-Venugopalan (MV) model~\cite{McLerran:1993ni, McLerran:1993ka}. Alternatively, parton saturation has also been studied by treating heavy-nucleus states as uncorrelated nucleon states~\cite{Mueller:1989st, Kovchegov:1996ty, Kovchegov:1998bi}. The latter approach shares some modeling characteristics with the Glauber model~\cite{Miller:2007ri}. The production of soft gluons at fixed order in the strong coupling $\alpha_s$ has been carried out in refs.~\cite{Kovchegov:2005ss, Wu:2017rry}, revealing qualitative differences from kinetic theory in the coupling expansion within $\phi^4$ theory~\cite{Kovchegov:2017way}. Recently, the broadening of high-energy partons in the early-stage color fields have been studied in ref.~\cite{Li:2021zaw, Avramescu:2023qvv}.

In this work, we investigate whether and how one may formulate hard processes and bulk matter on the same footing by modelling heavy-nuclei as uncorrelated nucleons. The focus will be on whether the impact-parameter dependent cross section, defined in ref.~\cite{Wu:2021ril}, factorizes at fixed order in $\alpha_s$ as well as at leading order in both the hard momentum scale and the nuclear size.
Instead of treating bulk matter and jets differently like in the previous studies, we carry out a detailed calculation of azimuthal decorrelation of the Drell-Yan (DY) and boson-jet processes in heavy-ion collisions. Such an observable has been extensively studied  in proton-proton~\cite{Mueller:1981fe, Collins:1984kg, Becher:2010tm, Buffing:2018ggv, Sun:2018icb, Chien:2019gyf, Chien:2020hzh, Chien:2022wiq} and high-energy nuclear collisions~\cite{Mueller:2016gko, Mueller:2016xoc, Chen:2016vem, Chen:2016cof, Chen:2018fqu, Arleo:2020rbm, Benic:2022ixp, Wang:2021jgm}. We consider the rescattering of the partons entering and leaving the hard scattering with partons from different nucleon-nucleon collisions. We do not include radiation.

The manuscript is organized as follows:
In Sec.~\ref{SecII} we define the collision parameter-dependent cross section in heavy ion collisions, the associated observable of interest in this study, namely the azimuthal decorrelation of the final state pair, which is either a Drell-Yan pair or a photon-jet pair.
In Sec.~\ref{SecIII} we define the Feynman rules used in momentum space and in coordinate space.
In Sec.~\ref{sec:coldonpdfs} we implement the cold nuclear effects in our observables and discuss their impact on the initial state distributions. These are all the modifications necessary to discuss the Drell-Yan pair azimuthal decorrelation.
In Sec.~\ref{SecV} we implement the cold nuclear effects for the hadronic final states, which are necessary to complete the discussion of the photon-jet azimuthal decorrelation. Finally, we discuss the form of the factorization obtained in this study.

\section{A unified description of hard processes and bulk matter}
\label{SecII}

In this section, we present the main formula to calculate the impact-parameter dependent cross section for hard processes in heavy-ion collisions. The constituent nucleons of the colliding nuclei are taken to be uncorrelated, following the perturbative approach to parton saturation in refs.~\cite{Mueller:1989st, Kovchegov:1998bi} and the Glauber model~\cite{Miller:2007ri}.

\subsection{The impact-parameter dependent cross section in heavy-ion collisions}\label{sec:dsigmadb}

In quantum theory, the impact-parameter dependent cross section for any observable $O$ in the collision of two ultra-relativistic particles can be generically defined as~\cite{Wu:2021ril}
\begin{align}\label{eq:dsigmadb}
        \frac{d\sigma}{d^2{\mathbf b} dO}     =&\int\prod\limits_f\left[d\Gamma_{p_f}\right]\delta(O-O(\{p_f\})\langle \phi_{1}\phi_{2}|\hat{S}^\dagger|\{ p_f\}\rangle\langle \{ p_f\}|\hat{S}|\phi_{1}\phi_{2}\rangle\notag\\
        =&\int\prod\limits_f\left[d\Gamma_{p_f}\right]\delta(O-O(\{p_f\})\text{Tr}\big[\hat{S}^\dagger|\{ p_f\}\rangle\langle \{ p_f\}|\hat{S}|\phi_{1}\phi_{2}\rangle\langle \phi_{1}\phi_{2}|\big],
\end{align}
where the wave packages of the colliding particles, denoted as $\phi_i$, are required to be sufficiently localized in space in order to define the impact parameter $\mathbf{b}$, and the phase-space measure for a particle with mass $m$ is defined as
\begin{align}\label{eq:Gammapf}
    \int d\Gamma_{p}\equiv \int \frac{d^4p}{(2\pi)^4}(2\pi)\delta(p^2-m^2)\theta(p^0).
\end{align}
Note that the unity in $\hat{S}$ should be disregarded, meaning $\hat{S}=1+i\hat{T}$ is replaced with $i\hat{T}$ to align with the conventional cross section definition. Here and below, boldface letters denote two-dimensional transverse vectors.

To calculate the impact-parameter dependent cross section in heavy-ion collisions using perturbative QCD, it is essential to have detailed information about the multi-parton distributions within nuclei. As such information is unavailable, below we express the nuclear multi-parton distributions in terms of those for their constituent nucleons by employing expansions based on physical scales.

The first relevant scale is the binding energy per nucleon, denoted as $\Delta E_{b}$. In the rest frame of heavy nuclei,  $\Delta E_{b}$ is known to be of the order of 10~MeV. This value provides an estimate of the typical reaction time between nucleons $t_n\sim 1/\Delta E_{b}\approx20$~fm/$c$. In the lab frame of ultra-relativistic heavy-ion collisions, $t_n$ is further dilated. Since $t_n$ is significantly longer than any other timescales, the interaction and correlation between constituent nucleons in the same nucleus will be neglected.

The second relevant length scale is the nuclear size $R$, which scales as $A^{1/3}$ with $A$ the mass number. As the kinetic energy per nucleon is of the same order as $\Delta E_b$, individual nucleons typically possess a momentum $\Delta p \sim \sqrt{2m_n \Delta E_b}\sim 140$ MeV, where $m_n$ denotes the nucleon mass. Consequently, the nucleons are localized in space with an uncertainty comparable to the nucleon size $\sim 1/\Lambda_{QCD}$. In the large $R$ limit, one can conduct an expansion in terms of $1/(R\Lambda_{QCD})$.  Consequently, the size of individual nucleons can be neglected at leading order in this expansion. Moreover, nucleons can be viewed as classical particles because $\Delta p$ is also negligible compared to other relevant momentum scales. The above expansions are, in essence, identical to the modelling in~\cite{Mueller:1989st, Kovchegov:1998bi, Miller:2007ri}.

Below, we assume that nucleus 1 with mass number $A_1$ moves along the positive $z$-axis, while nucleus 2 with mass number $A_2$ moves along the negative $z$-axis. In this case, the density matrix for nucleus 1 can be written in the form\footnote{For brevity, we do not distinguish neutrons from protons and omit all spin indices.}
\begin{align}
    |\phi_1\rangle\langle\phi_1| &= \prod\limits_{i=1}^{A_1}\int d\Gamma_{p_i} d\Gamma_{p'_i} | p_i\rangle\langle p_i|\phi_1\rangle\langle\phi_{1}| p'_i\rangle\langle p'_i|=\prod\limits_{i=1}^{A_1}\int\frac{dP_i^+d^2\mathbf{P}_i}{(2\pi)^32P_i^+}\notag\\ &\times\frac{1}{2}\int\,db_i^- d^2\mathbf{b}_i W_{A_1}(P_i, b_i)\int\frac{dq_i^+d^2\mathbf{q}_i}{(2\pi)^3} e^{\frac{i}{2}q_i^+ b_i^- - i \mathbf{q}_i\cdot\mathbf{b}_i}| P_i+{q_i}/{2}\rangle\langle P_i-{q_i}/{2}|,
\end{align}
where in the final expression on the right-hand side, we have replaced $1/p_i^+$ and $1/{p_i'}^{+}$ in $d\Gamma_{p_i} d\Gamma_{p'_i}$ by $1/P_i^+$, the Wigner distribution function for one nucleon  is defined as
\begin{align}
    W_{A_1}(P, b)\equiv \int\frac{dq^+d^2\mathbf{q}}{(2\pi)^3 2P^+}e^{-\frac{i}{2}q^+ b^- + i \mathbf{q}\cdot\mathbf{b}}\langle P+{q}/{2}|\phi_1\rangle\langle\phi_1| P-{q}/{2}\rangle,
\end{align}
and the light-cone components of a four-vector $v^\mu$ read
\begin{align}
    v^\pm = v^0\pm v^3.
\end{align}
As argued above, the constituent nucleons can be viewed as classical particles with~\cite{Kovchegov:2013cva, Wu:2017rry}
\begin{align}\label{eq:Wpb}
    W_{A_1}(p, b)=\hat{\rho}_{A_1}(b^-, \mathbf{b})2(2\pi)^3\delta(p^+-P_1^+)\delta^{(2)}(\mathbf{p}),
\end{align}
where $P^+_1$ denotes the "+" momentum of the nucleon, and $\hat{\rho}_{A_1}=\rho_{A_1}/A_1$ with $\rho_{A_1}(b^-, \mathbf{b})$ the distribution of nucleons (usually with a Woods-Saxon functional form). For nucleus 2, one only needs to swap the "+" and "$-$" variables and replace 1 by 2 in the above formulae.

\begin{figure}
    \centering
    \includegraphics[width=0.4\textwidth]{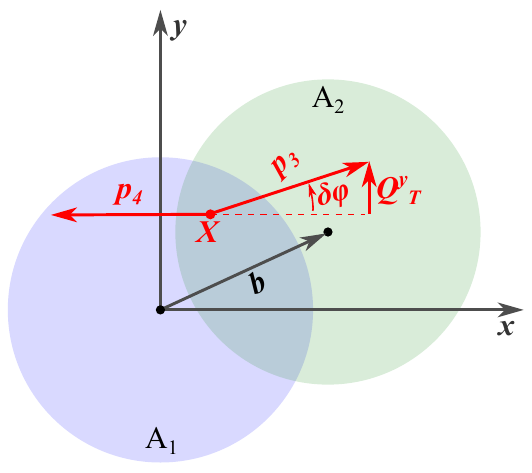}
    \caption{
    Azimuthal decorrelation in heavy-ion collisions. In the frame shown in this figure, the two colliding nuclei locate respectively at $\mathbf{0}$ and $\mathbf{b}$ in the transverse plane. For the two produced objects, the one with momentum $p_4$ is moving along the negative $x$-axis while the other one with momentum $p_3$ is moving predominantly along the positive $x$-axis. The azimuthal decorrelation $\delta\varphi$ is determined by the projection of $p_3$ onto the $y$-axis, denoted by $Q_T^y$.
    }
    \label{fig:obs}
\end{figure}

Expressing the density matrices in terms of the Wigner functions for both colliding nuclei in eq.~(\ref{eq:dsigmadb}) yields
\begin{align}\label{eq:dsigmadbJet}
        \frac{d\sigma}{d^2{\mathbf b} dO}     =&\int\prod\limits_f\left[d\Gamma_{p_f}\right]\delta(O-O(\{p_f\})\notag\\
        &\times\prod\limits_{i=1}^{A_1}\frac{1}{2P_{1}}\int\, d^2\mathbf{b}_i\,db_i^-\hat{\rho}_{A_1}(b_i^-, \mathbf{b}_i)\int\frac{dq_i^+d^2\mathbf{q}_i}{(2\pi)^3} e^{\frac{i}{2}q_i^+ b_i^-\,- i \mathbf{q}_i\cdot\mathbf{b}_i}
        \notag\\
        &\times\prod\limits_{j=1}^{A_2}\frac{1}{2P_{2}}\int\, d^2\mathbf{b}'_j\,db'^{+}_j\hat{\rho}_{A_2}(b'^{+}_j,\mathbf{b}'_j-\mathbf{b})\int\frac{dq_j^{\prime -}d^2\mathbf{q}'_j}{(2\pi)^3} e^{\frac{i}{2}q_j'^-\,b_j'^+\,- i \mathbf{q}'_j\cdot\mathbf{b}'_j}
        \notag\\
        &\times\langle \{P_{1}-{q_i}/{2}\}, \{P_{2}-{q'_j}/{2}\}|\hat{S}^\dagger|\{ p_f\}\rangle\langle \{ p_f\}|\hat{S}|\{P_{1}+{q_i}/{2}\}, \{P_{2}+{q'_j}/{2}\}\rangle,
\end{align}
where the momenta of the constituent nucleons within the two nuclei are respectively given by $P_{1}^{\mu}=P_{1}^+ n_1^{\mu}/2$ and  $P_{2}^{\mu}=P_{2}^- n_2^{\mu}/2$, with the two beam directions defined as $n_1^\mu =(1,0 , 0, 1)$ and $n_2^\mu=(1,0 , 0, -1)$. The impact-parameter dependent cross section in the above form clearly respects translation invariance. As illustrated in fig.~\ref{fig:obs}, we choose a frame in which nuclei 1 and 2 are located respectively at $\mathbf{0}$ and $\mathbf{b}$ in the transverse plane.

In the Glauber model~\cite{Miller:2007ri}, the observable $O$ represents multiplicity or energy deposition in a specific region of the detector. Since the impact parameter is correlated with this observable, collision geometry (centrality) is hence determined within the uncertainties dictated by fluctuations. In this work, we study the impact-dependent cross section for hard processes with soft particles all integrated over, as discussed in detail in ref.~\cite{Wu:2021ril}.

\subsection{The observable: azimuthal decorrelation}

To illustrate the utility of eq.~(\ref{eq:dsigmadbJet}),  we carry out a detailed calculation of the leading-order corrections to the azimuthal decorrelation $\delta\varphi$ between two final-state objects produced in a hard process due to the presence of nuclear matter. Such an observable has been proposed to study the properties of QCD matter in high-energy nuclear collisions~\cite{Mueller:2016gko, Mueller:2016xoc, Chen:2016vem, Chen:2016cof, Chen:2018fqu, Arleo:2020rbm, Benic:2022ixp, Wang:2021jgm}. In heavy-ion collisions, the hard process and bulk matter, conventionally studied separately using different approaches~\cite{Cao:2020wlm}, are coupled via phenomenological modelling. Here, we focus on how to formulate a unified description of both hard and soft sectors of the entire collision.

The definition of azimuthal decorrelation is illustrated in fig.~{\ref{fig:obs}}. Here,
the momenta for the incoming nuclei and the two final-state objects are represented  by $(A_1 P_{1}, A_2 P_{2})\to (p_3, p_4)$, respectively. The azimutal decorrelation $\delta\phi$ is determined by the component of $\mathbf{p}_3$ perpendicular to $\mathbf{p}_4$, denoted as $Q_T^y$, according to
\begin{align}
    \delta\varphi=\arcsin(Q_T^y/|\mathbf{p}_3|)
\end{align}
in the range $(-\pi, \pi)$. We focus on the kinematic region with $\delta\varphi\approx Q^y_T/Q \ll 1$, with the hard scale $Q = p_T\equiv|\mathbf{p}_4|\approx|\mathbf{p}_3| \gg |Q_T^y|$ in the Drell-Yan and boson-jet processes. To justify perturbative calculations, $Q_T^y$ is also taken to be semi-hard with  $|Q_T^y|\gg \Lambda_{QCD}$. In this back-to-back limit at hadron colliders, it is well-known that the azimuthal distribution is subject to Sudakov suppression as a result of soft and collinear radiation in the Drell-Yan~\cite{Mueller:1981fe, Collins:1985ue, Becher:2010tm} and boson-jet~\cite{Buffing:2018ggv, Sun:2018icb, Chien:2019gyf, Chien:2020hzh, Chien:2022wiq} processes. Below, we focus on additional corrections at tree level due to the presence of nuclear matter.

By exploiting rotation symmetry of the entire collision system in the transverse plane, the cross section for $\mathbf{b}$ and $\mathbf{p}_4$ at arbitrary orientations can be derived from
\begin{align}
\frac{1}{p_T}\frac{d\sigma_{A_1 A_2}}{d^2\mathbf{b} d\eta_3 d\eta_4 dp_T d\delta\varphi}\approx \frac{d\sigma_{A_1 A_2}}{d^2\mathbf{b} d\eta_3 d\eta_4 dp_T dQ_T^y}
\end{align}
with one of their orientations fixed. Here, $\eta_i$ is the pseudorapidity of final-state object $i$. We choose the frame depicted in fig.~\ref{fig:obs}, where $\mathbf{p}_4$ is aligned with the negative $x$-axis, to carry out our calculations.

At lowest order, there is no imbalance and the cross section for the underlying partonic process $i(p_1) + j(p_2)\to k(p_3) + l(p_4)$ is given by
\begin{align}\label{eq:sighat0th}
    \frac{d\hat\sigma^{(0)}_{ij\to kl}}{d\eta_3 d\eta_4 dp_T^2 d Q_T^y} = & \frac{1}{16\pi \hat{s}} \frac{\delta(Q_T^y)}{1+\delta_{kl}}|{\overline{M}^{(0)}_{ij\to kl}(p_1,p_2\to p_3, p_4)}|^2\notag\\
    &\times\delta(p_1^+-p_3^+-p_4^+)\delta(p_2^--p_3^--p_4^-),
\end{align}
where the bar over the square of the lowest-order amplitude $M^{(0)}_{ij\to kl}$ represents the average and sum over the initial- and final-state spins and colors respectively, $\hat{s}=(p_1+p_2)^2=(p_3+p_4)^2$, and $\delta_{kl}$ accounts for the statistical factor for identical particles. The partonic process involves the following four light-like vectors:
\begin{align}\label{eq:ns}
    &n_1^\mu = (1,0,0,1),\qquad n_2^\mu =(1,0,0,-1)\notag\\
    &n_k^{\mu}=(1, \sin\theta_k, 0, \cos\theta_k) = (1, 1/\cosh\eta_k, 0, \tanh\eta_k)
\end{align}
where $\theta_k$ denotes the polar angle of particle $k$ with $k=3, 4$. The four directions are assumed to be distinct, satisfying $n_i\cdot n_j\sim 1$ for $i\neq j$ with $i,j = 1,\dots, 4$.  In the following discussion, the lowest order expression in eq.~(\ref{eq:sighat0th}) is referred to as the zeroth order, and we shall evaluate the leading-order (LO) corrections that are enhanced by the medium length in the Drell-Yan and boson-jet processes.

\subsection{Perturbative description of hard processes and bulk matter}

\begin{figure}
    \centering
    \includegraphics[width=0.5\textwidth]{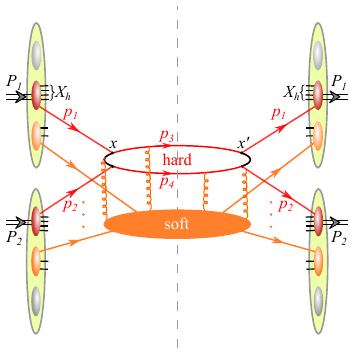}
    \caption{Unified description of hard processes and bulk matter. The uncorrelated nucleons are categorized into three groups: the ones initiating the hard processes (red), the ones participating the semi-hard (soft) process (orange) and spectators (gray).}
    \label{fig:genericGraphs}
\end{figure}

For azimuthal decorrelation there are two perturbative scales: the hard scale $Q$ and the soft scale $Q_T^y$. As all other final-state particles are integrated out, the cross section for such an observable is, in principle, amenable to perturbative calculations. As demonstrated in fig.~\ref{fig:genericGraphs}, a generic diagram can be divided into two coupled parts: the hard process characterized by $Q$, and the soft part characterized by $Q_T$. Even if the correlation between constituent nucleons is neglected, we do not know a priori whether the cross section in eq.~(\ref{eq:dsigmadbJet}) factorizes in the same manner as the corresponding processes in cleaner types of collisions like those summarized in~\cite{Collins:1989gx, Sterman:1995fz, Stewart:2003eft, Becher:2014oda, Kovchegov:2012mbw}. Therefore, we restrict ourselves to carrying out the leading-order corrections to the zeroth-order cross section in this work. 

Let us now express the cross section in terms of the parton distribution functions (PDFs) of the constituent nucleons. Following the parton model, we first define the nucleon PDFs using free bare fields in the interaction picture, which will be renormalized order by order in $\alpha_s$ whenever necessary. It is convenient to work in coordinate space. As illustrated in fig.~\ref{fig:genericGraphs}, one can take a pair of fields from the vertices located at $x$ and $x'$ in the amplitude and the conjugate amplitude, respectively. After acting the fields on one of the nucleon states in eq.~(\ref{eq:dsigmadbJet}), one has
\begin{align}
    &\int\, d^2\mathbf{b}\,db^{-}\hat{\rho}_{A_1}(b^{-}, \mathbf{b})\int\frac{dq^+d^2\mathbf{q}}{(2\pi)^3} e^{\frac{i}{2}q^{+} b^{-}\,- i \mathbf{q}\cdot\mathbf{b}}\notag\\
    &\times\sum\limits_{X_h}\langle P_{1} - {q}/{2}| \bar{\psi}^a_{\alpha}(x')|X_h\rangle\langle X_h|\psi^b_\beta(x) |P_{1} + {q}/{2}\rangle\notag\\
    &=\int\, d^2\mathbf{b}\,db^{-}\hat{\rho}_{A_1}(b^{-}, \mathbf{b})\int\frac{dq^+d^2\mathbf{q}}{(2\pi)^3} e^{\frac{i}{2}q^{+} (b^{-}-X^-)\,- i \mathbf{q}\cdot(\mathbf{b}-\mathbf{X})}\notag\\
    &\times\sum\limits_{X_h}\langle P_{1} - {q}/{2}| \bar{\psi}^a_{\alpha}(0)|X_h\rangle\langle X_h|\psi^b_\beta(0) |P_{1} + {q}/{2}\rangle e^{\frac{i}{2}(P_{1}^+-p^+_{X_h}) \Delta x^-},
\end{align}
where we consider, as an example, quark fields that carry both color (in superscript) and spinor (in subscript) indices, $X_h$  represents the states of the remnants, $X\equiv (x+x')/2$ and $\Delta x\equiv x'-x$. In the last step, we  have used the relation $\phi(x)=e^{i\hat{p}\cdot x}\phi(0)e^{-i\hat{p}\cdot x}$ with $\phi$ denoting a quantum field and $\hat{p}$ the four-momentum operator, as well as the fact that the nucleon and the remnants with their total momentum denoted by $p_{X_h}$ are moving predominantly along the positive $z$-axis. 

As we expand in $1/(R\Lambda_{QCD})$, we can consistently neglect $q$ in the hadron state in the above equation. This is justified by the fact that $q$ scales as the inverse of the nucleus size. Consequently, we obtain
\begin{align}
    &2\hat{\rho}_{A_1}(X^{-}, \mathbf{X})\langle P_{1}| \bar{\psi}^a_{\alpha}(\Delta x^- \bar{n}_1/2)\psi^b_\beta(0) |P_{1}\rangle\notag\\
    &=2\hat{\rho}_{A_1}(X^{-}, \mathbf{X})\frac{\delta^{ab}}{2 N_c} P_{1}^+ \frac{\slashed{n}_{1\beta\alpha}}{2}\int d\xi f_{q}(\xi) e^{i\frac{\Delta x^-}{2}\xi P^+_{1}}\notag\\
    &=2\hat{\rho}_{A_1}(X^{-}, \mathbf{X})\int \frac{d\xi}{\xi}f_{q}(\xi)  \frac{\delta^{ab} }{2 N_c}\slashed{p}_1 e^{i p_1\cdot(x'-x)},
\end{align}
where $\bar{n}_1^\mu\equiv(1,-\vec{n}_1)=n_2^\mu$ and $p^\mu_1=\xi P^\mu_{1}$, and $f_i$ represents the PDF for parton $i$ within the nucleon. Expanded at the same order, the quark distribution is consistent with the definition in QCD~\cite{Collins:1981uw}, which is given by
\begin{align}
    f_{q}(\xi)=\int \frac{dt}{2\pi}e^{-it \xi P_1^+}\langle P|\bar{\psi}(t\bar{n}_1)\mathcal{P}e^{-ig\int_0^t dt' \bar{n}_1\cdot A(t'\bar{n}_1)}\frac{\slashed{\bar{n}}_1}{2}\psi(0)|P\rangle
\end{align}
with $\mathcal{P}$ denoting the path-ordering operator. Applying the above reasoning to all fields acting on the hadron states within both nuclei, we finally obtain
\begin{align}\label{eq:dsigmadbJetFinal}
        \frac{d\sigma}{d^2{\mathbf b} dO}     =&\int\prod\limits_f\left[d\Gamma_{p_f}\right]\delta(O-O(\{p_f\})\notag\\
        &\times\sum\limits_{\{a_i, b_j\}}\bigg(\prod\limits_{i=1}^{A_1}\frac{1}{P_{1}^+}\int\frac{d\xi_i}{\xi_i} f_{a_i}(\xi_i)\bigg)
        \bigg(\prod\limits_{j=1}^{A_2}\frac{1}{P_{2}^-}\int\frac{d\xi'_j}{\xi'_j} f_{b_j}(\xi'_j)\bigg)\notag\\
        &\times\langle \{\xi_i P_{1}\}, \{\xi'_j P_{2}\}|\hat{S}^\dagger|\{ p_f\}\rangle\langle \{ p_f\}|\hat{S}|\{ \xi_i P_{1} \}, \{ \xi'_j P_{2}\}\rangle\notag\\
        &\otimes \bigg(\prod\limits_{i=1}^{A_1} \hat{\rho}_{A_1}(X^{-}_i, \mathbf{X}_i)\bigg)
        \bigg(\prod\limits_{j=1}^{A_2} \hat{\rho}_{A_2}(Y^{+}_j, \mathbf{Y}_j-\mathbf{b})\bigg),
\end{align}
where $a_i$ and $b_j$ iterate over all the parton species, and the operator $\otimes$ indicates that the incoming partons $i$ and $j$ enter the diagrams in the amplitude (the conjugate amplitude) at $x_i$ ($x_i'$) and $y_j$ ($y'_j$) respectively with $X_i = (x_i + x_i')/2$ and $Y_j=(y_j + y'_j)/2$. Here, the initial- and final-state spins and colors are respectively averaged and summed over in the square of the partonic $S$-matrix element. 

Since each of the $A_i$ nucleons in nucleus $i$ has an equal chance to participate the hard collision, at zeroth order, one has
\begin{align}
        \frac{d\sigma^{(0)}}{d^2{\mathbf b} dO}     =&\int\prod\limits_f\left[d\Gamma_{p_f}\right]\delta(O-O(\{p_f\})\frac{1}{s_{NN}}\sum\limits_{ij}\int\frac{d\xi}{\xi}\frac{d\xi'}{\xi'} f_i(\xi)f_j(\xi')
        \notag\\
        &\times\langle \xi P_{1}, \xi' P_{2}|\hat{S}^\dagger|\{ p_f\}\rangle\langle \{ p_f\}|\hat{S}|\xi P_{1}, \xi' P_{2}\rangle\otimes {\rho}_{A_1}(X^{-}, \mathbf{X})  {\rho}_{A_2}(Y^{+}, \mathbf{Y} - \mathbf{b})
\end{align}
with $s_{NN}=P_1^+ P_2^-$. In the expansion at large $Q$,  offshell propagators shrink to one single spacetime point, and we get
\begin{align}\label{eq:dsigmadb0}
        \frac{d\sigma^{(0)}}{d^2{\mathbf b} dO}     
        =& 2 \int d^4X{\rho}_{A_1}(X^{-}, \mathbf{X})  {\rho}_{A_2}(X^{+}, \mathbf{X} -\mathbf{b} )  
        \frac{d{\sigma}^{(0)}_{nn\to kl}}{dO},
\end{align}
where the nucleon-nucleon cross section is defined as
\begin{align}
      \frac{d{\sigma}^{(0)}_{nn\to kl}}{dO}\equiv\sum\limits_{ij}\int{d\xi}{d\xi'} f_i(\xi) f_j(\xi')
        \frac{d\hat{\sigma}^{(0)}_{ij\to kl}}{dO}.
\end{align}

In eq.~(\ref{eq:dsigmadbJetFinal}), the PDFs are defined in terms of free fields, and infrared divergences arise at high orders in $\alpha_s$. In Deep Inelastic Scattering,  proton-proton and proton-nucleus collisions, it has been justified, to varying degrees of rigor~\cite{Collins:1984kg, Collins:1989gx, Sterman:1995fz, Stewart:2003eft, Becher:2014oda, Kovchegov:2012mbw}, that infrared divergences arsing beyond LO in $\alpha_s$ can all be absorbed into universal parton distributions or jet functions defined by corresponding gauge invariant operators. Accordingly, the cross section in such collisions factorizes. In the subsequent sections, we shall examine whether the cross section in eq.~(\ref{eq:dsigmadbJetFinal}) for the Drell-Yan and boson-jet processes in heavy-ion collisions factorizes at fixed order.

\section{Expansion of Feynman diagrams at high $Q$}\label{sec:FeynmanRules}
\label{SecIII}

All our calculations in the following sections are to be carried out according to eq.~(\ref{eq:dsigmadbJetFinal}). As we are interested in the kinematic region with $\delta\varphi\ll 1$, we expand all the diagrams in $\delta\varphi$ and keep only leading-order terms in this expansion. With $Q_T^y$ held fixed, the diagrams are equivalently expanded at large $Q$. Furthermore, to be consistent with the expansion in the nucleus radius as discussed in sec.~\ref{sec:dsigmadb}, only contributions enhanced by the medium size will be kept among the leading-$Q$ terms. We present the Feynman rules based on this expansion below.

\subsection{Expansion in momentum space}

To organize the expansion of Feynman diagrams, let us first examine the scaling of momenta in a generic diagram, as illustrated in fig.~\ref{fig:genericGraphs}. There are up to four light-like directions defined by the four light-like vectors $n_i$.  At leading order, one only needs to consider the contributions to the azimuthal distribution due to momentum transfer $\sim \delta\varphi Q$ between different collinear directions.

The collinear partons engaged in the hard collision, upon interaction with the soft part of the diagrams, maintain their collinear directions. Consider a parton $i$ along the light-like direction given by $n_i$. The momentum of this $n_i$-collinear parton, denoted as $p$, scales as follows. It satisfies $n_j\cdot p \sim Q$ for $j \neq i$ as all the four light-like directions are assumed to be distinct. As we look for contributions enhanced by the medium size, collinear internal momenta should be nearly on-shell in order to travel for a long distance, so $2 n_i\cdot p \,n_j\cdot p/n_i\cdot n_j-|\mathbf{p}_\perp|^2\simeq 0$. That is, the $n_i$-collinear momentum $p$ scales as
\begin{align}
\label{eq:scalings}
    n_i\cdot p\sim \delta\varphi^2  Q,\qquad n_j\cdot p \sim Q\qquad\text{for $j \neq i$},
\end{align}
given the momentum transfer perpendicular to $n_i$ and $n_j$, denoted above by $\mathbf{p}_\perp$, scales as $\delta\varphi Q$. In the expansion at high $Q$, one can replace the propagator for a quark or antiquark with
\begin{align}\label{eq:SF}
S_F(p)=\slashed{p} D_F(p) \to \text{sign}(p^0)\slashed{p}_i D_F(p),
\end{align}
where the internal momentum $p$ is assumed to be aligned with the direction of fermion-number flow such that $\text{sign}(p^0)=1$ for quarks and $\text{sign}(p^0)=-1$ for antiquarks, $p_i$ is one of the incoming or outgoing momenta engaging in the hard collision and
\begin{align}
    D_F(p)\equiv\frac{i}{p^2+i\epsilon}.
\end{align}

The collinear partons couple to the soft part of the diagrams  through the exchange of gluons, represented by $A^{a\mu}$.
The Feynman rules for the interaction between collinear partons and the exchanged gluon fields have been widely employed in saturation physics~\cite{Kovchegov:2012mbw} and jet quenching physics~\cite{Casalderrey-Solana:2007knd}. Our emphasis here is on extending these rules to processes that involve multiple distinct collinear directions using the Feynman gauge. For an incoming $n_i$-collinear quark, upon expansion at high $Q$, one has,
\begin{align}\label{eq:qgqbEik}
\begin{array}{l}
\includegraphics[width=0.15\textwidth]{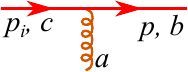}
\end{array}
&=-ig\slashed{p}_i D_F(p)\slashed{A} u_{p_i}= -2ig (p_i\cdot\, A)D_F(p)u_{p_i}\,,
\end{align}
with $A^{\mu}=A^{a\mu}t^a$, $t^a$ being the generators in the fundamental representation of $SU(N_c)$. Similarly, for an incoming antiquark, one only needs to replace $t^a_{bc}$ with $\bar{t}^a_{bc}=-t^{a}_{cb}$ in the above equation, along with a modification of the spinor. Here, the minus sign comes from the sign of $p^0$ in eq.~(\ref{eq:SF}). The above rules also apply to outgoing quarks and antiquarks with the corresponding spinors substitution.

The main complication, in comparison to the rules in physical gauges, arises from the interaction between collinear gluons and the exchanged gluons.
For an incoming gluon, in the Feynman gauge, one has
\begin{align}\label{eq:gggEik}
    \begin{array}{l}
     \includegraphics[width=0.2\textwidth]{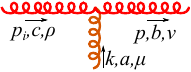}
    \end{array}
&=g f^{abc}A^{a}_{\mu}[g^{\mu\nu}(k+p)^{\rho} + g^{\nu\rho} (-p-p_i)^{\mu} + g^{\rho\mu}(p_i-k)^{\nu}]\epsilon_{\rho}(p_i)D_F(p)\notag\\
&\approx -gf^{abc}[2 p_i\cdot A^a \epsilon^{\nu}(p)  - A^a\cdot\epsilon(p)~p_i^{\nu}]D_F(p).
\end{align}
Given that physical polarization vectors $\epsilon^\mu$ are used for all the incoming or outgoing gluons, the second term, contracted with the rest part of the diagrams, vanishes (at leading order in $Q$) according to the Ward identity (see, e.g., \cite{Collins:1989gx})
\begin{align}
    \langle M|T\partial_{\mu_1} A^{\mu_1}(x_1)\cdots \partial_{\mu_n} A^{\mu_n}(x_n)|N\rangle=0,
\end{align}
given that $M$ and $N$ are physical states. That is, one can safely take
\begin{align} 
\label{eq:ggg}
\begin{array}{l}
     \includegraphics[width=0.2\textwidth]{image/ggg}
    \end{array}
&= -2gf^{abc}(p_i\cdot A^a)\epsilon^{\nu}(p_i)D_F(p)
= -ig T^{a}_{bc} 2 p_i \cdot A^a ~\epsilon^{\nu}(p_i)D_F(p),
\end{align}
in the expansion at large $Q$. This rule also applies to outgoing collinear gluons.

As illustrated in fig.~\ref{fig:genericGraphs}, the soft part of the diagrams can also couple to the hard part by attaching $A^{a\mu}$ onto the internal propagator in the hard process. However,  these diagrams can be effectively eliminated through an expansion in $Q$, as such an insertion only splits the off-shell propagator into two off-shell propagators. Since neither of the resulting propagators are close to being on-shell, these diagrams are not enhanced by the medium size, in contrast with the insertions of the exchanged gluon onto the external legs.

As a result, it is sufficient to consider diagrams where the soft part couples to the external legs of the hard part. As we see above, the spinors and polarization vectors of all the incoming and outgoing partons pass through the vertices on the external legs, contracting directly with the off-shell part of the diagrams. Then, expanding the off-shell part at high $Q$ yields the zeroth-order amplitudes $i M^{(0)}$. In this way, the medium corrections all factorize from the amplitude for the hard process on the amplitude level. One only needs to evaluate the interaction between the external legs and the gluon fields $A^{a\mu}$ produced by the soft part of the diagrams using the following rules:
\begin{align}\label{eq:FeynmanRules}
\begin{array}{l}
\includegraphics[width=0.15\textwidth]{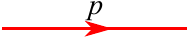}
\end{array}
= D_F(p),   
\qquad
\begin{array}{l}
\includegraphics[width=0.15\textwidth]{image/qqg}
\end{array}
= -ig \hat{T}^a 2p_i\cdot A^a,
\end{align}
where all collinear partons are represented by solid lines, and $\hat{T}^a$ stands for the color generator: $\hat{T}^a_{bc}=t^a_{bc}$ for quarks, $\hat{T}^a_{bc}=-t^a_{cb}$ for antiquarks, and $\hat{T}^a_{bc}=-if^{abc}$ for gluons.  

The rest of this work focuses on interactions between collinear partons via one gluon exchange. According to the Feynman rules in eq.~(\ref{eq:FeynmanRules}), the interaction between two collinear partons along the same direction is negligible at high $Q$. One only needs to consider the interaction between  partons collinear to different directions. Given two different light-cone directions $n_i$ and $n_j$, the momentum of the exchanged gluon scales as $n_i\cdot p \sim \delta\varphi^2 
 Q\sim n_j\cdot p$ according to momentum conservation. That is, one has $n_i\cdot p \,n_j\cdot p \sim \delta\varphi^4 
 Q^2$, which can always be expanded. Therefore, for a so-called Glauber gluon~\cite{Bodwin:1981fv} connecting two collinear partons along $n_i$ and $n_j$ respectively, one has
\begin{align}\label{eq:GGlauberp}
\begin{array}{l}
\includegraphics[width=0.15\textwidth]{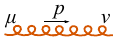}
\end{array}
=  \frac{i g_{\mu\nu}}{|\mathbf{p}_\perp|^2}
 \end{align}
 with $|\mathbf{p}_\perp|^2 = - g_\perp^{\mu\nu} p_{\mu} p_{
 \nu
}$
and the transverse metric tensor defined as
\begin{align}
    g^{\mu\nu}_\perp=g^{\mu\nu}-\frac{n_i^{\mu}n_j^{\nu} + n_i^{\nu}n_j^{\mu}}{n_{ij}}\qquad\text{with $n_{ij}\equiv\,n_i\cdot\,n_j$}.
\end{align}
 Here and below, boldface letters with a subscript $\perp$ denote two-dimensional vectors perpendicular to two light-like directions $n_i$ and $n_j$. To be consistent with previous conventions, the subscript is omitted if $n_i$ and $n_j$ are both the two beam directions. That is, the subscript $\perp$ is reserved as a reminder that at least one of the final-state collinear partons is involved in the interaction.

For the two beam directions $n_1$ and $n_2$, the transverse $x$ and $y$ components of a vector are defined as illustrated in fig.~\ref{fig:obs}. When $\vec{n}_i$ is not parallel to $\vec{n}_j$, we define the transverse "1" and "2" components using the following two basis vectors:
\begin{align}\label{eq:nPerps}
    n_{\perp_1}^\mu = \frac{1}{\sqrt{1-(\vec{n}_{i}\cdot \vec{n}_{j})^2}}\bigg(1+\vec{n}_i\cdot\vec{n}_j, \vec{n}_i + \vec{n}_j\bigg),\qquad n_{\perp_2}^\mu = \bigg(0, \frac{\vec{n}_i\times \vec{n}_j}{|\vec{n}_i\times \vec{n}_j|}\bigg).
\end{align}
As the four three-vectors $\vec{n}_i$ are in the same plane, $n_{\perp_2}^\mu$ is universal (up to an overall minus sign) for any pair of $n_i$ and $n_j$, and it is chosen to be aligned with the $y$-axis. Accordingly, a four-vector $v^\mu$ can be decomposed as
\begin{align}\label{eq:LCcoor}
    v^\mu = \frac{v^{n_i} n_j^{\mu} + v^{n_j} n_i^{\mu}}{n_{ij}} + v^1  n_{\perp_1}^\mu + v^2 n_{\perp_2}^\mu\ ,
\end{align}
with
\begin{align}    v^{n_i}\equiv\,n_i\cdot\,v,\qquad\,v^{n_j}\equiv\,n_j\cdot\,v
,
\end{align}
and the transverse vector $\mathbf{v}_\perp=(v^1, v^2)$ with $|\mathbf{v}_{\perp}|^2 = (v^1)^2 + (v^2)^2$.

\subsection{Expansion in coordinate space}
\label{sec:FeynmanrulesCoor}
It turns out to be more convenient to work in coordinate space. As the amplitude for the hard process only depends on $Q$ and, hence, factorizes in the high-$Q$ expansion, all the off-shell propagators shrink into one single spacetime point. Now, one only needs to Fourier transform the building blocks for the external legs in eqs.~(\ref{eq:FeynmanRules}) and~(\ref{eq:GGlauberp}) into coordinate space. Or equivalently, the $S$-matrix elements in eq.~(\ref{eq:dsigmadbJetFinal}) are calculated directly in coordinate space: the corresponding parton-level diagrams are evaluated using the vertices given above in eq.~(\ref{eq:FeynmanRules}) and the free propagators given below in eqs.~(\ref{eq:DFFinal}) and (\ref{eq:DFGlauber}), convoluted with
\begin{align}
\begin{array}{l}
\includegraphics[width=0.25\textwidth]{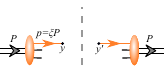}
\end{array}
=\frac{1}{\bar{n}\cdot P}\int\frac{d\xi}{\xi} f_{i}(\xi)\hat{\rho}_{A}(n\cdot(y+y')/2, (\mathbf{y}+\mathbf{y}')/2-\mathbf{b})
\end{align}
for each incoming or outgoing collinear parton moving along the $n$ direction. Here, $\bar{n}^{\mu}=(1,-\vec{n})$ and $\mathbf{b}$ stands for the transverse location of nucleus $A$. All the incoming (outgoing) collinear partons converge at (originate from) a single spacetime point representing the hard process, with the coefficient given by $iM^{(0)}$. Note that $iM^{(0)}$ only depends on the on-shell momenta of the incoming and outgoing particles.

 The Feynman rules above are justified as follows. As the momentum transfer is always between two collinear directions, in the presence of another collinear direction $n_j$, an internal momentum $p$ collinear to the $n_i$-direction scales as $p_\perp\sim \delta\varphi Q,$ $n_i\cdot p\sim \delta\varphi^2 Q$, $ n_j\cdot p\sim Q$ for $j\neq i$, see eq.~(\ref{eq:scalings}).
In coordinate space, one has
\begin{align}
    x_\perp\sim 1/(\delta\varphi Q),\qquad n_i\cdot x\sim 1/Q, \qquad n_j\cdot x\sim 1/(\delta\varphi^2 Q).
\end{align}

For an $n_i$-collinear propagator coupled to another $n_j$-collinear parton, we decompose $x^{\mu} = (x^{n_i}, x^{n_j}, x_\perp)$ according to eq.~(\ref{eq:LCcoor}). From the scalar product $x\cdot p = \frac{1}{n_{ij}} ( x^{n_i} p^{n_j} + x^{n_j} p^{n_i} ) - x^1 p^1 - x^2 p^2$, one can see that the metric tensor in this coordinate system is given by
\begin{align}
    g_{\mu\nu} = \left(
    \begin{array}{cccc}
       0  & \frac{1}{n_{ij}} & 0 & 0 \\
       \frac{1}{n_{ij}} & 0 & 0 & 0 \\
       0 & 0 & -1 & 0\\
       0 & 0 & 0 & -1
    \end{array}
    \right).
\end{align}
With this metric, the Feynman propagator $D_F$ in coordinate space can be expressed as
\begin{align}\label{eq:DF}
    D_F(x)&=\int\frac{d^4p}{(2\pi)^4}\sqrt{|g|}\frac{i e^{-ip\cdot x}}{p^2+i\epsilon}=\frac{1}{n_{ij}}\int\frac{dp^{n_i} dp^{n_j}d^2\mathbf{p}_{\perp}}{(2\pi)^4}\frac{ie^{-ip\cdot x}}{p^2+i\epsilon}\notag\\
    &= \int\frac{dp^{n_j}}{2\pi}\frac{1}{2p^{n_j}} e^{-\frac{i}{n_{ij}}p^{n_j} x^{n_i}}G_0(x^{n_j}, \mathbf{x}_{\perp};p^{n_j})[\theta(p^{n_j})\theta(x^{n_j})-\theta(-p^{n_j})\theta(-x^{n_j})],
\end{align}
where the propagator for a non-relativistic particle in two dimensions is defined as
\begin{align}
    G_0(x^{n_j}, \mathbf{x}_{\perp} ; p^{n_j})\equiv\int\frac{d^2\mathbf{p}_{\perp}}{(2\pi)^2}e^{-\frac{i}{n_{ij}} \frac{|\mathbf{p}_{\perp}|^2}{p^{n_j}} x^{n_j}+i\mathbf{p}_{\perp}\cdot\mathbf{x}_{\perp}} = \frac{n_{ij} p^{n_j}}{4\pi i x^{n_j}} e^{\frac{i}{2} \frac{n_{ij}}{2} \frac{p^{n_j}}{x^{n_j}} |\mathbf{x}_\perp|^2}.
\end{align}
As all the coordinate variables are eventually integrated over in our calculations, $x$ can be so chosen that the corresponding initial- (final-) state parton flows toward (away from) the hard vertex. In this case, the $\theta$ functions in eq.~(\ref{eq:DF}) can be dropped irrespective of parton species.

In the equation above, $x^{n_j}$ plays the role of time and $p^{n_j}$ corresponds to the mass (sometimes referred to as Galilean mass in the context of Light-Cone Perturbation Theory~\cite{Beuf:2016wdz,Kogut:1969xa}). That is, the $n_j$-component (resp.~$n_i$-component) is an analog of the "+" component (resp.~"$-$" component) in light-cone coordinates. 
We have not specified the relation between $1/p^{n_i}\sim 1/(\delta^2\varphi Q)$ and the length of the QCD medium. For simplicity, we assume the path length to be much smaller than $1/(\delta^2\varphi Q)$ in the following discussions, yielding
\begin{align}\label{eq:DFFinal}
    D_F(x)=\int\frac{dp^{n_j}}{2\pi}\frac{1}{2p^{n_j}} e^{-\frac{i}{n_{ij}}p^{n_j} x^{n_i}} \delta^{(2)}(\mathbf{x}_\perp)\theta(x^{n_j})
\end{align}
with $p^{n_j}>0$ as per the convention mentioned above.

In the coordinates describe before, the Glauber gluon propagator connecting the two collinear directions $n_i$ and $n_j$ in eq.~(\ref{eq:GGlauberp}) takes the form
\begin{align}\label{eq:DFGlauber}
    G_F^{\mu\nu}(x)& = ig^{\mu\nu}  n_{ij} \delta(x^{n_i})\delta(x^{n_j})F_1(|\mathbf{x}_{\perp}|),
\end{align}
where in dimensional regularization with $ d = 2 - 2\epsilon$, $F_1$ is defined as~\cite{Li:2023vdj}
\begin{align}
F_1(|\mathbf{x}|)&\equiv\mu^{2-d}\int\frac{d^d \mathbf{l}}{(2\pi)^d} \frac{e^{i\mathbf{l}\cdot \mathbf{x}}}{|\mathbf{l}|^2}=\frac{1}{4\pi}\frac{\Gamma(-\epsilon)}{(\pi |\mathbf{x}|^2 \mu^2)^{-\epsilon}}=-\frac{1}{4\pi}\bigg[\frac{1}{\epsilon} +\gamma_E+\ln (\pi |\mathbf{x}|^2 \mu^2)+O(\epsilon)\bigg]. 
\end{align}
Here, the infrared divergence results from replacing nucleon states by parton states within them in eq.~(\ref{eq:dsigmadbJetFinal}). It reflects the fact that the PDFs in that equation need to be renormalized. Otherwise, the infrared divergence can be regularized by modelling nucleons, e.g., with color-singlet dipoles~\cite{Mueller:1989st, Kovchegov:1996ty, Li:2023vdj}.

\section{Cold nuclear effects on spatial parton distributions}
\label{sec:coldonpdfs}

In this section, we study the azimuthal decorrelation in the Drell-Yan process due to the presence of QCD matter. We first carry out a detailed calculation of $O(\alpha_s)$ corrections due to the interaction of the incoming partons engaged in the hard collision with the nucleons they encounter. Then, we proceed with the resummation of these terms in an expansion on the nuclear size.

\subsection{Initial-state single scattering}

\subsubsection{The single-scattering diagrams}

\begin{figure}
    \centering
    \includegraphics[width= \textwidth]{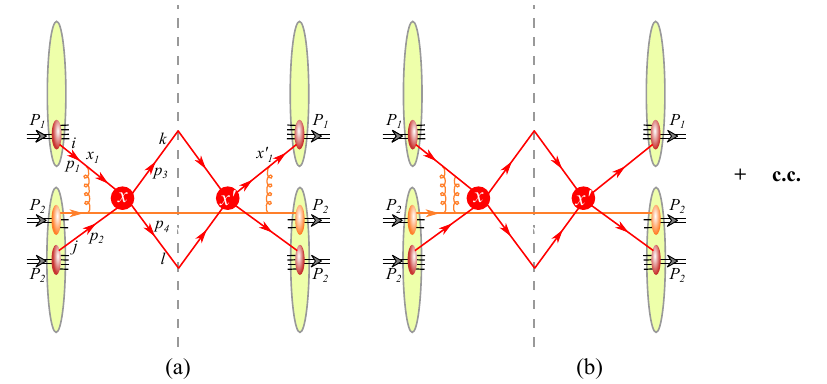}
    \caption{Cold nuclear effects before the hard collision associated with $i(p_1)+j(p_2)\to k(p_3)+l(p_4)$. 
    The left panel show the real diagram for single scattering in which one incoming parton exchanges one Glauber gluon with one encounter nucleon in both the amplitude and the conjugate amplitude. The right panel shows the corresponding virtual diagrams in which the incoming parton does not interact with the nucleon either in the amplitude or the conjugate amplitude.}
    \label{fig:LO_before}
\end{figure}

In ultra-relativistic heavy-ion collisions, the longitudinal sizes of colliding nuclei experience substantial Lorentz contraction. However, in contrast to soft gluon production, the hard collision involving high-$x$ partons occurs within a distance even smaller than the contracted sizes of the nuclei. With the increase in nuclear density compensating for the contracted path length, the likelihood of an incoming parton interacting with nucleons from the oppositely moving nucleus remains the same as that in DIS, current-nucleus and nucleon-nucleus collisions~\cite{Kovchegov:1998bi}.

Let us start with diagrams in which one additional nucleon couples to the two incoming partons before they engage in the hard collision. The additional nucleon can be either the constituent of nucleus 1 or 2. The incoming partons in eq.~(\ref{eq:dsigmadbJetFinal}) all carry some intrinsic transverse momenta of the order of $\Lambda_{QCD}$, which are negligible in comparison with $\delta\varphi Q$. Therefore, one can expect that it is negligible for the hard process to generate the measured $Q_T^y$ by simply absorbing one parton from the addiontal nucleon. In this case, the leading-order correction due to nuclear effects starts at $O(\alpha_s^2)$ in the fixed-order expansion.

Among all the diagrams at $O(\alpha_s^2)$,  it is sufficient to  include those where each incoming hard parton interacts with the nucleon moving in the opposite direction. Consequently, there are only two types of real diagrams to be considered. Moreover, only the $t$-channel exchange graphs between the incoming partons and the additional nucleons need to be kept, as any $s$- or $u$-channel exchange would put the ensuing external leg offshell. After squaring these diagrams, the additional nucleons have to be the same in both the amplitude and in the conjugate amplitude to ensure color neutrality, as exemplified by the diagrams in fig.~\ref{fig:LO_before}. In this figure, the corresponding virtual diagrams are also included to guarantee unitarity.

The parton from the additional nucleon, after exchanging one gluon with the hard process, is integrated out.  Let us single out this part of diagrams, as shown in fig.~\ref{fig:LO_before}~(a). Without loss of generality, we take the parton from the additional nucleon to be an valence quark in the following discussion. Assuming the valence quark moving along the negative-$z$ axis, this part of the diagram reads
\begin{align}
\label{eq:AA}
&\begin{array}{l}
     \includegraphics[width=0.24\textwidth]{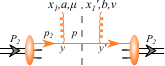}
\end{array}
\equiv
\frac{\delta^{ab}}{N_c^2-1}\langle A^{c\mu}(x_1)A^{c\nu}(x'_1)\rangle
= \frac{A_2 -1}{A_2} \frac{1}{P_2^-}\int \frac{d\xi}{\xi} f_q(\xi)\int\,d\Gamma_p\\
&\times\frac{\text{Tr}(t^a t^b)}{2 N_c}g^2\text{Tr}(\slashed{p}\gamma_\alpha\slashed{p}_2\gamma_{\beta})\int d^4y d^4y'\rho_{A_2}(Y^+, \mathbf{Y} - \mathbf{b})e^{i\Delta y
\cdot(p-p_2)}G_F^{\mu\alpha}(x_1-y)G_F^{\nu\beta}(x_1'-y')\notag
\end{align}
with $p_2=\xi P_2$, $Y=(y+y')/2$ and $\Delta y = y - y'$. Here, the factor $(A_2-1)/A_2$ arises from the fact that one out of the $A_2$ nucleons participates in the hard process. This factor is taken to be unity below, as we only consider cases with $A_{1,2}\gg 1$. We then expand away high order terms in $\delta\varphi$ in the parton-level amplitude according to the scaling: $p^-\sim Q$, $|\mathbf{p}|\sim \delta\varphi$ and $p^+\sim \delta\varphi^2 Q$ and drop $p^+$ everywhere in the above equation. After using the expression for the Glauber gluon propagator in eq.~(\ref{eq:DFGlauber}), one can easily obtain
\begin{align}
\label{eq:AAbefore}
\langle A^{c\mu}(x_1)A^{c\nu}(x'_1)\rangle
=&
g^2 C_F n_2^{\mu}n_2^{\nu} \delta( x_1^+ - {x_1'}^+)\notag\\
&\times\int d^2\mathbf{y} \rho_{A_2}(x_1^+, \mathbf{y} - \mathbf{b})\int d\xi f_q(\xi)F_1(|\mathbf{x}_1-\mathbf{y}|)F_1(|\mathbf{x}'_1-\mathbf{y}|).
\end{align}
It is equivalently given by the product of the classical fields in ref.~\cite{Kovchegov:1996ty}. That is, the valence quark behaves like a classical color charge moving along the light cone, which is identical to that in the MV model~\cite{McLerran:1993ni, McLerran:1993ka}.

We first evaluate the real diagram squared in fig.~\ref{fig:LO_before}~(a), using eq.~(\ref{eq:dsigmadbJetFinal}) along with Feynman rules in coordinate space as presented in sec.~\ref{sec:FeynmanrulesCoor}. In this case, the incoming hard parton $i$ is moving along $n_i=n_1$ while the additional nucleon is moving along $n_j=n_2$. Accordingly, we use the light-cone variables: $v^{n_i}=v^-$ and $v^{n_j}=v^+$. In terms of the above field correlator, one has
\begin{align}
    \text{fig.~\ref{fig:LO_before}~(a)}=&\frac{2g^2 C_
i  }{N_c^2-1}\int d\xi d\xi'\int d^4 x d^4x' d^4 x_1 d^4 x'_1 (p_1^+)^2\hat{H}_{ij\to kl}(X_1^-, \mathbf{X}_1, \xi; X^+, \mathbf{X}, \xi')\notag\\
&\times \langle A^{a-}(x_1) A^{a-}(x_1')\rangle D_F(x-x_1) D_F^*(x'-x_1') e^{-i(x-x')\cdot(p_2-p_3-p_4) - i (x_1-x'_1)\cdot p_1},
\end{align}
where $X\equiv (x+x')/2$, $X_1\equiv (x_1+x_1')/2$, $C_i$ is the quadratic Casimir corresponding to the color representation of parton $i$, and
\begin{align}\label{eq:singleBefore}
    \hat{H}_{ij\to kl} \equiv& \frac{f_i(\xi) f_j(\xi')}{2\xi\xi's_{NN}} \rho_{A_1}(X_1^{-}, \mathbf{X}_1)
    \rho_{A_2}(X^{+}, \mathbf{X}-\mathbf{b}) |\overline{M}^{(0)}_{ij\to kl}(p_1, p_2; p_3, p_4)|^2
\end{align}
with $s_{NN}=(P_1+P_2)^2 = P_1^+ P_2^-$, $p_1=\xi P_{1}$ and $p_2=\xi' P_{2}$. One can then integrate over $(x^-, \mathbf{x})$ and $(x'^-, \mathbf{x}')$ to obtain
\begin{align}
    \text{fig.~\ref{fig:LO_before}~(a)}=&\frac{g^2 C_
i}{2(N_c^2-1)}\int dx^{+} d{x'}^{+} d^4 x_1 d^4 x'_1\int d\xi d\xi'  \hat{H}_{ij\to kl}(X^-, \mathbf{X}, \xi; X^+, \mathbf{X}, \xi')\notag\\
&\times \langle A^{a-}(x_1) A^{a-}(x_1')\rangle  \theta( x^+ - x^+_1) \theta( {x'}^+_1- {x_1'}^+) \bigg(\frac{p_1^+}{p_3^+ + p_4^+}\bigg)^2
\notag\\
&\times e^{\frac{i}{2}(x^+ - {x'}^+)(p_3^- + p_4^- - p_2^-) - i(\mathbf{x}_1 - \mathbf{x}'_1 )\cdot \mathbf{Q}_T + \frac{i}{2} (x_1^- - {x'}_1^-) (p_3^+ + p_4^+ - p_1^+)},
\end{align}
where $\mathbf{Q}_T \equiv \mathbf{p}_3 + \mathbf{p}_4$, and in light-cone coordinates $X^\mu \equiv( (x^+ + {x'}^+)/2, (x_1^- + {x'}_1^-)/2, (\mathbf{x}_1 + \mathbf{x}_1')/2)$ can be viewed as the spacetime point where the hard collision takes place.

The above result applies to all field correlators. To proceed, one can make use of the properties of the correlator in eq.~(\ref{eq:AAbefore}), specifically its proportionality to $\delta(x_1^+ - {x'}_1^+)$ and independence of $x_1^-$ and ${x'}_1^-$. Consequently, by integrating over $x_1^- - {x'}_1^-$ and $x^+ - {x'}^+$, one obtains
\begin{align}\label{eq:sigma0Forsingle}
    \text{fig.~\ref{fig:LO_before}~(a)}=&\frac{g^2 C_
i}{N_c^2-1}\int d^4X \rho_{A_1}(X^{-}, \mathbf{X})
    \rho_{A_2}(X^{+}, \mathbf{X} - \mathbf{b})
    \int_{-\infty}^{X^+} d X_1^+
    d\sigma^{(0)}_{ij\to kl}(X^+-X_1^+) \notag\\
&\times \int  d t d^2\mathbf{x} e^{- i\mathbf{x}\cdot \mathbf{Q}_T}
\langle A^{a-}(X_1^+ + t, \mathbf{X} + {\mathbf{x}}/{2}) A^{a-}(X_1^+ - t, \mathbf{X} - {\mathbf{x}}/{2})\rangle,
\end{align}
where $t\equiv (x_1^+ - {x'}_1^+)/2$, $\mathbf{x}$ is redefined as $\mathbf{x}_1 - \mathbf{x}'_1$ in eq.~(\ref{eq:singleBefore}), and the zeroth-order differential cross section is given by
\begin{align}\label{eq:sig0_LO_before}
    d\sigma^{(0)}_{ij\to kl}(X^+)\equiv& 
   \int \frac{d\xi}{\xi} \frac{d\xi'}{\xi'} \frac{1}{2 s_{NN}} f_{i}(\xi)
    f_{j}(\xi')|\overline{M}^{(0)}_{ij\to kl}(p_1, p_2; p_3, p_4)|^2\notag\\
   &\times(4\pi)\delta( p_1^+ - p_3^+ - p_4^+)\frac{2\sin[X^+(p_2^- - p_3^- - p_4^-)]}{p_2^- - p_3^- - p_4^-}.
\end{align}
Note that  the "$-$" momentum is not conserved exactly in $d\sigma^{(0)}_{ij\to kl}(X, X_1^+)$, in contrast to the full result at the zeroth order in eq.~(\ref{eq:dsigmadb0}). This is a consequence of the uncertainty principle as the nucleon are confined in the nucleus. However, this should not be taken literally as we have already expanded away higher-order terms in $1/(RQ)$.

Similarly, for the virtual diagrams in fig.~\ref{fig:LO_before}~(b), one can first evaluate the required field correlator to obtain
\begin{align}
\begin{array}{l}
\begin{tikzpicture}
    \node at (0,0) {\includegraphics[width=0.5\textwidth]{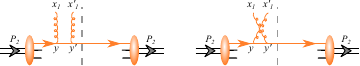}};
    \node[below] at (0,0) {+};
\end{tikzpicture}
\end{array}
=&
-\frac{\delta^{ab}}{N_c^2-1}\langle A^{c\mu}(x_1)A^{c\nu}(x'_1)\rangle,
\end{align}
where we have used the relation
$
\lim\limits_{{x'}_1^+\to x_1^+}[\theta(x_1^+ - {x'}_1^+)+\theta({x'}_1^+ - x_1^+)]=1$.
At the end, including the contribution from virtual diagrams as illustrated in fig.~\ref{fig:LO_before}~(b) yields\footnote{This result has been obtained by imposing unitarity.  We find that it is necessary to require the Glauber exchange to occur prior to the "$+$" time of the hard process in both the amplitude and the conjugate amplitude for the virtual diagrams.}
\begin{align}\label{eq:singleDY}
    \text{fig.~\ref{fig:LO_before}}=& - \frac{g^2 C_
i}{2(N_c^2-1)}\int d^4X \rho_{A_1}(X^{-}, \mathbf{X}) 
    \rho_{A_2}(X^{+}, \mathbf{X} - \mathbf{b})
    \int_{-\infty}^{X^+} d X_1^+
    d\sigma^{(0)}_{ij\to kl}(X^+-X_1^+) \notag\\
&\times\int  d t d^2\mathbf{x}  e^{- i\mathbf{x}\cdot \mathbf{Q}_T}
\langle [A^{a-}(X_1^+ + t, \mathbf{X} + {\mathbf{x}}/{2}) - A^{a-}(X_1^+ - t, \mathbf{X} - {\mathbf{x}}/{2})]^2\rangle.
\end{align}
The contributions from the diagrams in which the incoming parton $j$ interacts with an additional nucleon from nucleus 1 can be obtained by replacing $C_i$ with $C_j$ and swapping the indices of nuclei 1 and 2 as well as the "+"  and "-" variables.

\subsubsection{The phase-space gluon distributions in heavy nuclei}

In hadron-hadron collisions, the only dominate source for a collinear parton to acquire a transverse momentum of the order of $\delta\varphi Q\gg \Lambda_{QCD}$ is radiation. Such contributions can be systematically organized based on Transverse Momentum Dependent (TMD) factorization~\cite{Collins:1984kg, Becher:2010tm, Buffing:2018ggv, Sun:2018icb, Chien:2019gyf, Chien:2020hzh, Chien:2022wiq}. In the presence of other nucleons in heavy-ion collisions, the physical picture revealed by the result in eq.~(\ref{eq:singleDY}) is qualitatively different: each incoming hard parton scatters with the phase-space gluon distributions in the oppositely-moving nucleus before participating in the hard collision.

In order to substantiate this interpretation, let us evaluate
\begin{align}\label{eq:GAABefore}
    G&(X_1^+, \mathbf{X}, \mathbf{x})\equiv\int  d t 
\langle [A^{a-}(X_1^+ + t, \mathbf{X} + {\mathbf{x}}/{2}) - A^{a-}(X_1^+ - t, \mathbf{X} - {\mathbf{x}}/{2})]^2\rangle\notag\\
&
=2 g^2 C_F \int d^2\mathbf{r}\rho_{A_2}(X_1^+, \mathbf{X} - \mathbf{r})\int d\xi f_{q}(\xi)[F_1(|\mathbf{r} + \mathbf{x}/2|) - F_1(|\mathbf{r} - \mathbf{x}/2|)]^2\notag\\
&=  \frac{\alpha_s C_F}{2\pi}\int d^2\mathbf{r} \rho_{A_2}(X_1^+, \mathbf{X} - \mathbf{r})\int d\xi f_q(\xi) \ln^2\bigg(\frac{|\mathbf{r} + \mathbf{x}/2|^2}{|\mathbf{r} - \mathbf{x}/2|^2}\bigg),
\end{align}
where the vector $\mathbf{r} \equiv \mathbf{X} - \mathbf{y}$ is the transverse distance vector between the two scattering partons and the impact parameter $\mathbf{b}$ is omitted. One can easily see that the above integral contains a large logarithm with an non-perturbative infrared cutoff of $O(1/R)$, as prescribed by $\rho_{A_2}$. Such a large logarithm needs to be absorbed into the PDFs, and we accordingly introduce a factorization scale $\mu$ to renormalize the PDFs.\footnote{
The factorization scale $\mu$ should be of order $\Lambda_{QCD}$, which could be set in by modelling the nucleons by color singlet dipoles~\cite{Mueller:1989st, Kovchegov:1996ty, Li:2023vdj}}
For the logarithmic term, one can expand $\rho_{A_2}$ around $\mathbf{r}=0$. In this expansion, one has
\begin{align}\label{eq:thicknessBeam}
    G&(X_1^+, \mathbf{X}, \mathbf{x})=\pi |\mathbf{x}|^2 \rho_{A_2}(X_1^+, \mathbf{X}) \lim\limits_{x\to 0}xG(x, 1/|\mathbf{x}|^2) + \text{nonlogarithmic terms},
\end{align}
where the gluon distribution generated by the quark in a nucleon is given by
\begin{align}
    xG(x, 1/|\mathbf{x}|^2) = \frac{\alpha_s C_F}{\pi}\ln\bigg(\frac{1}{\mu^2 |\mathbf{x}|^2}\bigg)\int d\xi f_q(\xi, \mu).
\end{align}
Although it coincides with the solution of the DGLAP equation at $\mathcal{O}(\alpha_s)$ for $x\ll1$, it represents the Fourier transform of the transverse-momentum dependent gluon distribution under the same approximation. Consequently, one can recognize the Fourier transform of the phase-space gluon distribution, or the thickness beam function, within the nucleus as defined in ref.~\cite{Wu:2021ril}
\begin{align}
    xf_{g/A_2}(X_1^+, \mathbf{X}, x, \mathbf{x}) =  \rho_{A_2}(X_1^+, \mathbf{X})xG(x, 1/|\mathbf{x}|^2)
\end{align}
at $x\ll1$.

\subsubsection{Azimuthal decorrelation due to single scattering in the Drell-Yan process}

Let us insert the logarithmic term in eq.~(\ref{eq:thicknessBeam}) into the result of the single scattering in eq.~(\ref{eq:singleDY}) and specify the observable $O$ as $\mathbf{Q}_T$ at given $p_T$, $\eta_3$ and $\eta_4$ in eq.~(\ref{eq:dsigmadbJetFinal}). In terms of the zeroth-order cross section: 
\begin{align}\label{eq:dsig0_LO_before}
\frac{d\sigma^{(0)}_{ij\to l^+ l^-}}{ d\eta_3 d\eta_4 dp_T}(X^+)\equiv \frac{p_T}{4(2\pi)^3}d\sigma^{(0)}_{ij\to l^+ l^-}(X^+)
\end{align}
with $d\sigma^{(0)}_{ij\to l^+ l^-}(X^+)$ given in eq.~(\ref{eq:sig0_LO_before}) for fig.~\ref{fig:LO_before},
incorporating additional contributions by swapping nuclei 1 and 2 finally yields
\begin{align}\label{eq:ResultFigIII}
    \frac{d\sigma^{(1)}_{A_1 A_2\to l^+ l^-}}{d^2\mathbf{b} d\eta_3 d\eta_4 dp_T d^2\mathbf{Q}_T} =&-\frac{1}{2}\int d^4X \rho_{A_1}(X^{-}, \mathbf{X}) 
    \rho_{A_2}(X^{+}, \mathbf{X} - \mathbf{b}) \int \frac{d^2\mathbf{x}}{(2\pi)^2} e^{- i\mathbf{x}\cdot \mathbf{Q}_T} |\mathbf{x}|^2\notag\\
\times&\sum\limits_{ij}\bigg[\int_{-\infty}^{X^+} d X_1^+\frac{d\sigma^{(0)}_{ij\to l^+ l^-}}{ d\eta_3 d\eta_4 dp_T}(X^+-X_1^+) \hat{q}_{i/A_2}(X_1^+, \mathbf{X} - \mathbf{b}, |\mathbf{x}|)\notag\\
&+\int_{-\infty}^{X^-} d X_1^-\frac{d\sigma^{(0)}_{ij\to l^+ l^-}}{ d\eta_3 d\eta_4 dp_T}(X^- - X_1^-) \hat{q}_{j/A_1}(X_1^-, \mathbf{X}, |\mathbf{x}|)\bigg],
\end{align}
where the "+" and "-" momenta are interchanged in the zeroth-order cross section in the last term on the right-hand side, and the jet quenching parameter is given by
\begin{align}\label{eq:JetQuenchingParameter}
    \hat{q}_{j/A_i}(X^\pm, \mathbf{X}, |\mathbf{x}|)=\frac{4\pi^2 \alpha_s C_
j}{N_c^2-1} \rho_{A_i}(X^\pm, \mathbf{X}) xG(x, 1/|\mathbf{x}|^2).
\end{align}
This parameter is the same as defined in ref.~\cite{Baier:1996sk}, except that the scale in the gluon distribution is dictated by the observable with $1/|\mathbf{x}|\sim Q_T$.

\subsection{Multiple scattering for the Drell-Yan process in large nuclei}

Let us now focus on large nuclei and restrict our consideration to leading-order corrections in the nuclear size $R$. In this case, for typical values of $X^+$ and $X_1^{+}$ in eq.~(\ref{eq:sigma0Forsingle}), one can make the following replacement
\begin{align}\label{eq:deltaApp}
    \frac{2\sin[(X^+-X_1^+)(p_2^- - p_3^- - p_4^-)]}{p_2^- - p_3^- - p_4^-}\to 2\pi\delta(p_2^- - p_3^- - p_4^-).
\end{align}
In this approximation, the zeroth-order nucleon-nucleon cross section in eq.~(\ref{eq:dsig0_LO_before}) is the same as that in the zeroth-order nucleus-nucleus collisions, as presented in eq.~(\ref{eq:dsigmadb0}). Combining the zeroth-order and the single-scattering results results in
\begin{align}
    \frac{d\sigma_{A_1 A_2\to l^+ l^-}}{d^2\mathbf{b} d\eta_3 d\eta_4 dp_T d^2\mathbf{Q}_T}
    =& 2 \int d^4X \rho_{A_1}(X^{-}, \mathbf{X}) 
    \rho_{A_2}(X^{+}, \mathbf{X} - \mathbf{b}) \sum\limits_{ij} \frac{d\sigma^{(0)}_{ij\to l^+ l^-}}{ d\eta_3 d\eta_4 dp_T}\notag\\
    \times& \int \frac{d^2\mathbf{x}}{(2\pi)^2} e^{- i\mathbf{x}\cdot \mathbf{Q}_T} \bigg[ 1 - 
    \frac{|\mathbf{x}|^2}{4}\int_{-\infty}^{X^+} d X_1^+ \hat{q}_{i/A_2}(X_1^+, \mathbf{X} - \mathbf{b}, |\mathbf{x}|)\notag\\
    &- \frac{|\mathbf{x}|^2}{4}\int_{-\infty}^{X^-} d X_1^- \hat{q}_{j/A_1}(X_1^-, \mathbf{X}, |\mathbf{x}|)\bigg].
\end{align}

As each incoming hard parton only interacts with oppositely-moving nucleons, the cold nuclear effects on the two incoming partons do not interfere. Therefore, one can straightforwardly iterate our calculations for single scattering to obtain the results for muliple scatterings. Given that the average number of scattering is significantly smaller than the mass numbers $A_i$, the summation of multiple scattering results can be substituted by the exponentiation of single scattering results, namely:
\begin{align}
    \frac{d\sigma_{A_1 A_2\to l^+ l^-}}{d^2\mathbf{b} d\eta_3 d\eta_4 dp_T d^2\mathbf{Q}_T}
    =& 2 \int d^4X \rho_{A_1}(X^{-}, \mathbf{X}) 
    \rho_{A_2}(X^{+}, \mathbf{X} - \mathbf{b}) \sum\limits_{ij} \frac{d\sigma^{(0)}_{ij\to l^+ l^-}}{ d\eta_3 d\eta_4 dp_T} \int \frac{d^2\mathbf{x}}{(2\pi)^2} \notag\\
    \times& e^{- i\mathbf{x}\cdot \mathbf{Q}_T - 
    \frac{|\mathbf{x}|^2}{4}\int_{-\infty}^{X^+} d X_1^+ \hat{q}_{i/A_2}(X_1^+, \mathbf{X} - \mathbf{b}, |\mathbf{x}|)- \frac{|\mathbf{x}|^2}{4}\int_{-\infty}^{X^-} d X_1^- \hat{q}_{j/A_1}(X_1^-, \mathbf{X}, |\mathbf{x}|)}.
\end{align}
That is, each incoming parton, before participating in the hard collision, experiences an equivalent amount of momentum broadening as an individual parton in QCD medium, as studied in refs.~\cite{Baier:1996sk, Kovchegov:1998bi}.

In \cite{Balitsky:2017gis,Balitsky:2023hmh,Balitsky:2024ozy}, higher twist corrections to the TMD involved in Drell-Yan were computed in a rapidity factorization framework. These corrections have the symbolic form
\begin{equation}\label{eq:higherBalitsky}
    \langle F^{+\perp}F^{+\perp} \rangle_A \langle F^{-\perp}F^{-\perp} \rangle_B + \langle F^{+\perp}F^{+\perp}F^{+\perp} \rangle_A \langle F^{-\perp}F^{-\perp}F^{-\perp} \rangle_B
\end{equation}
in the case of gluon TMD factorization. It is understood that Wilson lines in a Drell-Yan staple are involved in the above expression, but are omitted here for clarity, and the average is over the background fields generated by the plus-mover hadron $A$ or the minus-mover hadron $B$. The twist two TMDs are given by the first term, and the second term contains higher twist corrections.
In the present case we focus on the corrections enhanced by $A^{1/3}$. In the Glauber model, these would take the following symbolic form
\begin{equation}
      \langle F^{+\perp}F^{+\perp}F^{+\perp}F^{+\perp} \rangle_A \langle F^{-\perp}F^{-\perp}\rangle_B +   \langle F^{+\perp}F^{+\perp} \rangle_A \langle F^{-\perp}F^{-\perp}F^{-\perp}F^{-\perp}\rangle_B\,,
\end{equation}
which is consistent with \cite{Collins:2007nk,Qiu:2001hj}.
The higher twist contributions in eq.~(\ref{eq:higherBalitsky}) do not appear in our case. This simply follows from the fact that each nucleon is a color singlet and the nucleons are uncorrelated in the nucleus.
As a final remark about the Glauber-gluon cancellation usually observed at leading twist, the contribution enhanced by $A^{1/3}$ is a higher twist effect where the cancellation is only partial, which can be seen in, e.g., eq.~(\ref{eq:singleDY}) where the field average would not vanish between real and virtual contributions.

\section{Cold nuclear effects on jets}
\label{SecV}

In this section, we focus on the azimuthal decorrelation in boson-jet production due to the presence of QCD matter. The interactions with QCD medium before the hard collision is identical to the Drell-Yan process as studied in the previous section. Below, we evaluate the corrections after the hard collision and the interference between the interactions before and after the hard collision. 

\subsection{The coordinates for jet production}

For a jet collinear to $n_3^\mu = (1, \sin\theta_3, 0, \cos\theta_3)$, it is convenient to choose the four basis vectors according to eq.~(\ref{eq:LCcoor}) with $n_i=n_3$ and $n_j=n_1$ or $n_2$. When the jet interacts with right-moving nucleons, one can choose $n_j^\mu=n_1^\mu=(1,0,0,1)$, yielding
\begin{align}
    n_{\perp_1}^\mu = (n_3^+/\sin\theta_3, 1, 0, n_3^+/\sin\theta_3),\qquad
    n_{\perp_2}^\mu = (0, 0, 1, 0).
\end{align}
Accordingly, the Feynman propagator in eq.~(\ref{eq:DFFinal}) reads explicitly
\begin{align}\label{eq:DFn3n1}
    D_F(x)&= \int\frac{dp^{-}}{2\pi}\frac{1}{2p^{-}} e^{-\frac{i}{n_{3}^-}p^{-} x^{n_3}}\theta(x^{-})\delta^{(2)}( \mathbf{x}_{\perp}),
\end{align}
indicating that, in terms of $x^-$ and $\mathbf{x}_{\perp}$, the jet parton follows the trajectory of a classical particle moving in the $n_3$ direction. To be more specific, according to the decomposition in eq.~(\ref{eq:LCcoor}), the transverse components of $x^\mu$ with respect to the beam direction can be expressed as
\begin{align}\label{eq:nPerps31}
    \mathbf{x} = \frac{x^- \mathbf{n}_3 + x^{n_3}\mathbf{n}_1}{n_3^-} + x^1 \mathbf{n}_{\perp_1} + x^2 \mathbf{n}_{\perp_2} =\bigg(\frac{x^- \sin\theta_3}{n_3^-}, 0\bigg) + \mathbf{x}_\perp.
\end{align}
This expression is identical to the transverse components of the trajectory of a classical particle moving along $n_3$:
\begin{align}
    x^\mu(t) = (0, x^1, x^2, 0) + n_3^\mu t = \bigg(t, \frac{x^-(t)\sin\theta_3}{n_3^-}+x^1, x^2, t\cos\theta_3\bigg)
\end{align}
with $x^-(t) = t-v_z t=n_3^- t$.

Similarly, for the interaction between the jet and left-moving nucleons, we choose $n_j^\mu=n_2^\mu=(1,0,0,-1)$. Accordingly, the transverse basis vectors in eq.~(\ref{eq:nPerps}) take the form
\begin{align}\label{eq:nPerps32}
    n_{\perp_1}^\mu = (n_3^-/\sin\theta_3, 1, 0, -n_3^-/\sin\theta_3),\qquad
    n_{\perp_2}^\mu = (0, 0, 1, 0),
\end{align}
and
\begin{align}\label{eq:DFn3n2}
    D_F(x)&= \int\frac{dp^{+}}{2\pi}\frac{1}{2p^{+}} e^{-\frac{i}{n_{3}^+}p^{+} x^{n_3}}\theta(x^{+})\delta^{(2)}( \mathbf{x}_{\perp}).
\end{align}
This indicates that, in terms of $x^+$ and $\mathbf{x}_{\perp}$, the transverse location of the jet parton with respect to the beam directions can be expressed as
\begin{align}\label{eq:xTxperp}
    \mathbf{x} = \frac{x^+ \mathbf{n}_3 + x^{n_3}\mathbf{n}_2}{n_3^+} + x^1 \mathbf{n}_{\perp_1} + x^2 \mathbf{n}_{\perp_2} =\bigg(\frac{x^+ \sin\theta_3}{n_3^+},0\bigg) + \mathbf{x}_\perp,
\end{align}
which also coincides with the transverse trajectory of a classical particle moving along the $n_3$ direction.

\subsection{Single scattering of a jet in QCD medium}

\subsubsection{Single scattering after the hard collision}

\begin{figure}
    \centering
    \includegraphics[width= \textwidth]{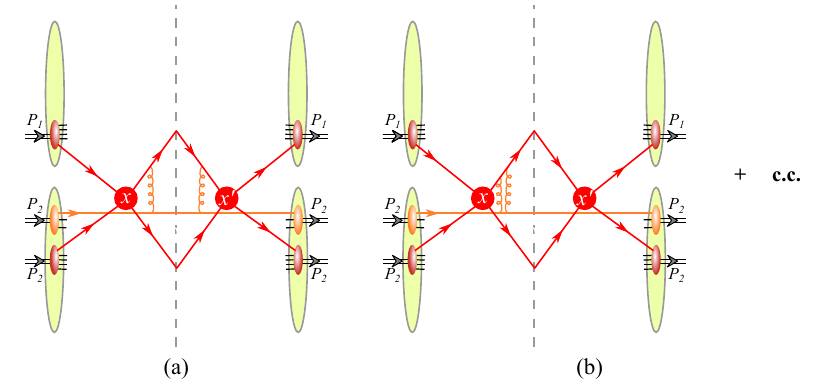}
    \caption{Cold nuclear effects after the hard collisions. The left panel show the real diagram for single scattering in which the outgoing parton exchanges one gluon with one encounter nucleon in both the amplitude and the conjugate amplitude. The right panel shows the corresponding virtual diagrams in which the outgoing parton does not interact with the nucleon either in the amplitude or the conjugate amplitude.}
    \label{fig:LO_after}
\end{figure}

The outgoing jet can interact with nucleons in either of the nuclei. Let us start with single scattering between the outgoing parton with one additonal nucleon from nucleus 2, as shown in fig. \ref{fig:LO_after}.

The field correlator is expressed in the same way as that before the hard collision, as given by eq.~(\ref{eq:AA}). Here, one only needs to evaluate it in the coordinate system defined by $n_2$, $n_3$, and the transverse basis vectors given in eq.~(\ref{eq:nPerps32}). After inserting $D_F$ in eq.~(\ref{eq:DFFinal}) and $G_F^{\mu\nu}$ in eq.~(\ref{eq:DFGlauber}) into eq.~(\ref{eq:AA}), we find
\begin{align}
\label{eq:AAAfter}
    \langle A^{b\mu}( x^+, \mathbf{x}_{\perp} )
    A^{b\mu'}(x'^+, \mathbf{x}'_{\perp} )\rangle=&
g^2 C_F  n_2^{\mu'}n_2^{\mu} \delta( x^+ - {x'}^+)\int d^2\mathbf{y}_{\perp}\rho_{A_2}(x^+, \mathbf{y} - \mathbf{b})\notag\\
\times&\int d\xi f_q(\xi)F_1(|\mathbf{x}_{\perp}-\mathbf{y}_{\perp}|)F_1(|\mathbf{x}'_{\perp}-\mathbf{y}_{\perp}|),
\end{align}
where the measure $d^2\mathbf{y}_{\perp}=dy^1 dy^2$, and $\mathbf{y}$, the transverse vector with respect to the beams in the nucleon density, is given by
\begin{align}\label{eq:yT}
    \mathbf{y} = \bigg( \frac{\sin\theta_3}{n_3^+} x^+ + y^1,  y^2\bigg)=\bigg( \frac{\sin\theta_3}{n_3^+} x^+,  0\bigg)+ \mathbf{y}_{\perp}
\end{align}
according to eq.~(\ref{eq:xTxperp}).
Note, for example, in terms of $( x^+, \mathbf{x}_{\perp} )$, one has $\mathbf{x} = ( \sin\theta_3 x^+/{n_3^+} + x^1,  x^2)$ and, consequently, $|\mathbf{x}_{\perp}-\mathbf{y}_{\perp}|=|\mathbf{x}-\mathbf{y}|$. This shows that the above basis vectors provide a convenient description of the interaction between the outgoing parton and nucleons close to its trajectory.

Following the evaluation of fig.~\ref{fig:LO_before}~(a) with the minus components of coordinates and momenta replaced with their corresponding $n_3$-components, one has
\begin{align}
    \text{fig.~\ref{fig:LO_after}~(a)}=&\frac{g^2 C_
k}{N_c^2-1}\int d^4X\int^{\infty}_{X^+} d X_1^+ \rho_{A_1}(X^{-}, \mathbf{X})
    \rho_{A_2}(X^{+}, \mathbf{X} - \mathbf{b}) d\sigma^{(0)}_{ij\to kl}(X_1^+-X^+) \notag\\
&\times\int  d t d^2\mathbf{x}_\perp e^{- i\mathbf{x}_{\perp}\cdot \mathbf{Q}_\perp}
\langle A^{a-}(X_1^+ + t, \mathbf{X}_{\perp} + {\mathbf{x}_{\perp}}/{2}) A^{a-}(X_1^+ - t, \mathbf{X}_\perp - {\mathbf{x}_\perp}/{2})\rangle ,
\end{align}
where $\mathbf{X}_\perp$ is given by
\begin{align}
\mathbf{X}_\perp = (-n_{\perp_1}\cdot X, -n_{\perp_2}\cdot X)=\mathbf{X} - ( \sin\theta_3 X^+/n_3^+, 0),
\end{align}
and 
\begin{align}
    \mathbf{Q}_\perp = (-n_{\perp_1}\cdot (p_3 + p_4 - p_1), Q_T^y) = (p_3^1, Q_T^y).    
\end{align}
Here, we have used the relation
\begin{align}
    p_3^{\mu} = \frac{p_T}{\sin\theta_3} n_3^{\mu} + \frac{\sin\theta_3}{2 p_T}|\mathbf{p}_{3\perp}|^2 n_2^{\mu}  + p_{3\perp}^{\mu}, \qquad p_4^{\mu} = p_T(\cosh\eta_4, -1, 0, \sinh\eta_4),
\end{align}
and, to obtain $d\sigma^{(0)}_{ij\to kl}$,
\begin{align}
    \frac{2\sin[\frac{2}{n_3^+}(X_1^+ - X^+)n_3\cdot(p_1 + p_2 - p_4)]}{\frac{2}{n_3^+}n_3\cdot(p_1 + p_2 - p_4)}
    =\frac{2\sin[(X_1^+ - X^+)(p_1^- + p_2^- - p_4^-)]}{p_1^- + p_2^- - p_4^-},
\end{align}
where $p_1^+ = p_3^+ + p_4^+$ and terms $\propto |\mathbf{p}_{3\perp}|^2/p_T$ are neglected.

Then, including the virtual diagrams as exemplified by fig.~\ref{fig:LO_after}~(b) gives
\begin{align}
    \text{fig.~\ref{fig:LO_after}}=&\frac{-g^2 C_
k}{2(N_c^2-1)}\int d^4X\int^{\infty}_{X^+} d X_1^+ \rho_{A_1}(X^{-}, \mathbf{X})
    \rho_{A_2}(X^{+}, \mathbf{X} - \mathbf{b}) d\sigma^{(0)}_{ij\to kl}(X_1^+-X^+)\int d^2\mathbf{x}_\perp  \notag\\
&\times\int  d t e^{- i\mathbf{x}_{\perp}\cdot \mathbf{Q}_\perp}
\langle [A^{a-}(X_1^+ + t, \mathbf{X}_{\perp} + {\mathbf{x}_{\perp}}/{2}) - A^{a-}(X_1^+ - t, \mathbf{X}_\perp - {\mathbf{x}_\perp}/{2})]^2\rangle.
\end{align}
The expectation value of the gluon field operator
\begin{align}
    G&(X_1^+, \mathbf{X}_\perp, \mathbf{x}_\perp)\equiv\int  d t 
\langle [A^{a-}(X_1^+ + t, \mathbf{X}_\perp + {\mathbf{x}_\perp}/{2}) - A^{a-}(X_1^+ - t, \mathbf{X}_\perp - {\mathbf{x}}_\perp/{2})]^2\rangle
\end{align}
also contains a logarithm in a non-perturbative scale, which needs to be renormalized. We then change variables from $\mathbf{y}_\perp$ in eq.~(\ref{eq:AAAfter}) to $\mathbf{r}_\perp = \mathbf{X}_\perp - \mathbf{y}_\perp$, and keep only the logarithmic term in $|\mathbf{x}_\perp|$ after integrating over $\mathbf{r}_\perp$ to obtain
\begin{align}
    G&(X_1^+, \mathbf{X}_\perp, \mathbf{x}_\perp)\approx\pi |\mathbf{x}_\perp|^2\rho_{A_2}(X_1^+, \mathbf{X}((X_1^+-X^+)/n_3^+)) xG(0, 1/|\mathbf{x}_\perp|^2),
\end{align}
where the transverse location of the jet at $t$ is defined as
\begin{align}
    \mathbf{X}(t) \equiv \mathbf{X} + {(\sin\theta_3, 0)}t,
\end{align}
which is the same as $\mathbf{y}$ in eq.~(\ref{eq:yT}) upon expansion at small $\mathbf{r}_\perp$:
\begin{align}
    \mathbf{y}=\bigg( \frac{\sin\theta_3}{n_3^+} X_1^+,  0\bigg)+ \mathbf{y}_{\perp}
    \approx
    \bigg( \frac{\sin\theta_3}{n_3^+} X_1^+,  0\bigg)+ \mathbf{X}_{\perp}=\bigg( \frac{\sin\theta_3}{n_3^+} (X_1^+-X^+),  0\bigg) + \mathbf{X}.
\end{align}
Accordingly, one has
\begin{align}\label{eq:ResultFigIV}
    \text{fig.~\ref{fig:LO_after}}=&-\frac{1}{2}\int d^4X\int^{\infty}_{X^+} d X_1^+ \rho_{A_1}(X^{-}, \mathbf{X})
    \rho_{A_2}(X^{+}, \mathbf{X} - \mathbf{b}) d\sigma^{(0)}_{ij\to kl}(X_1^+-X^+) \notag\\
&\times\int d^2\mathbf{x}_\perp e^{- i\mathbf{x}_{\perp}\cdot \mathbf{Q}_\perp} |\mathbf{x}_\perp|^2
\hat{q}_{k/A_2}(X_1^+, \mathbf{X}((X_1^+-X^+)/n_3^+) - \mathbf{b}, |\mathbf{x}_\perp|).
\end{align}

\subsubsection{Incoherence between initial- and final-state effects}

\begin{figure}
    \centering
    \includegraphics[width= \textwidth]{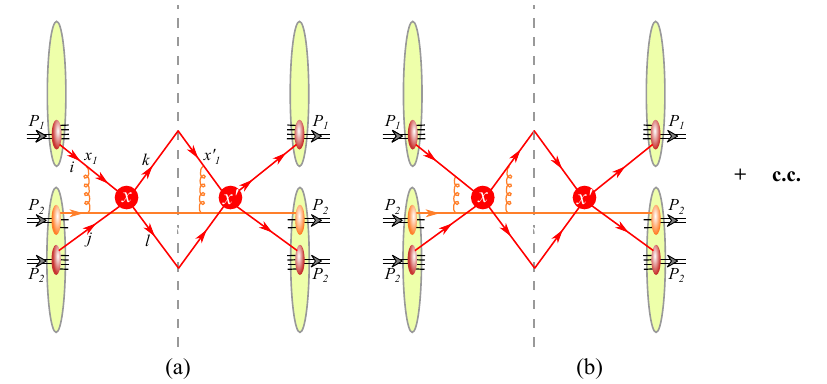}
    \caption{Interference between single scattering before and after the hard collision $ij\to kl$. The left panel show the real diagram for single scattering in which two different partons engaging in the hard collisions exchanges one gluon with one encounter nucleon in the amplitude and the conjugate amplitude. The right panel shows the corresponding virtual diagrams in which the incoming parton does not interact with the nucleon either in the amplitude or the conjugate amplitude.}
    \label{fig:LO_interfere}
\end{figure}

The interference diagrams between single collision before and after the hard collision, as shown in fig.~\ref{fig:LO_interfere}, are power suppressed. For virtual diagrams in fig.~\ref{fig:LO_interfere}~(b), the integration range for $x^+$ shrinks to zero. Consequently, such diagrams can be neglected. For a real diagram such as fig.~\ref{fig:LO_interfere}~(a), in order to reach the same conclusion as the virtual diagrams, we also needs to require ${x'}^+ > {x}_1^+$ and ${x'}_1^+ > x^+$. 
This feature is inherent to the time order perturbation theory. The hard process is defined as a localized event, which, in accordance with the uncertainty principle, is given by the inverse of the hard scale. Consequently, all other time scales involved are larger than this time scale, which allows us to identify both times $x_1^+$ and ${x'}_1^+$. Further discussions can be found in~\cite{Collins:1983ju} and in references therein.

\subsubsection{Azimuthal decorrelation due to single scattering in the boson-jet process}

Gathering all the above single scattering results, which include single scattering occurring before the hard collision, as summarized in eqs.~(\ref{eq:ResultFigIII}) for the Drell-Yan process, and single scattering occurring after the hard collision, as given in eq.~(\ref{eq:ResultFigIV}), yields
\begin{align}
    \frac{d\sigma^{(1)}_{A_1 A_2\to J V}}{d^2\mathbf{b} d\eta_3 d\eta_4 dp_T dQ_T^y} =&-\frac{1}{2}\int d^4X \rho_{A_1}(X^{-}, \mathbf{X}) 
    \rho_{A_2}(X^{+}, \mathbf{X} - \mathbf{b}) \int \frac{dy}{2\pi} e^{- i y Q^y_T} y^2\notag\\
\times\sum\limits_{ijk}\bigg[\int_{-\infty}^{X^+} d X_1^+ & \frac{d\sigma^{(0)}_{ij\to k V}}{ d\eta_3 d\eta_4 dp_T}(X^+-X_1^+) \hat{q}_{i/A_2}(X_1^+, \mathbf{X} - \mathbf{b}, |y|)\notag\\
+\int_{-\infty}^{X^-} d X_1^- & \frac{d\sigma^{(0)}_{ij\to  k V}}{ d\eta_3 d\eta_4 dp_T}(X^- - X_1^-) \hat{q}_{j/A_1}(X_1^-, \mathbf{X}, |y|)\notag\\
+\int^{\infty}_{X^+} d X_1^+ & \frac{d\sigma^{(0)}_{ij\to k V}}{ d\eta_3 d\eta_4 dp_T}(X_1^+ - X^+) \hat{q}_{k/A_2}(X_1^+, \mathbf{X}((X_1^+ - X^+)/n_3^+) - \mathbf{b}, |y|)\notag\\
+\int^{\infty}_{X^-} d X_1^- & \frac{d\sigma^{(0)}_{ij\to  k V}}{ d\eta_3 d\eta_4 dp_T}(X_1^- - X^-) \hat{q}_{k/A_1}(X_1^-, \mathbf{X}((X_1^- - X^-)/n_3^-), |y|)
\bigg].
\end{align}
Here, we have kept only the $y$-component of $\mathbf{Q}_T$ by integrating out $Q_T^x$ for single scattering that happens before the hard collision, and by integrating out $Q_T^1$ for single scattering that occurs after the hard collision.

\subsection{The boson-jet azimuthal decorrelation in large nuclei}
In the large $R$ expansion, it is reasonable to make the same replacement as eq.~(\ref{eq:deltaApp}) in $d\sigma^{(0)}_{ij\to kl}$. In this case, one can sum over the results for multiple scattering and finally obtains
\begin{align}\label{eq:kV_sum}
    &\frac{d\sigma_{A_1 A_2\to JV}}{d^2\mathbf{b}d\eta_3 d\eta_4 d p_T d Q_T^y} =\sum\limits_{ijk}  \frac{d\sigma^{(0)}_{ij\to kV}}{d\eta_3 d\eta_4 d p_T } \int dX^+ dX^- d^2\mathbf{X} {\rho}_{A_1}(X^{-}, \mathbf{X} ) \, {\rho}_{A_2}(X^{+}, \mathbf{X}-\mathbf{b})\notag\\
    &~~~~\times \int \frac{d y}{2\pi}  e^{- i y Q^y_T - \frac{y^2}{4}\int_{-\infty}^{X^+} d X_1^{+}\hat{q}_{i/A_2}(X_1^+, \mathbf{X} - \mathbf{b}, |y|) - \frac{y^2}{4}\int_{-\infty}^{X^-} d X_1^{-}\hat{q}_{j/A_1}(X_1^-, \mathbf{X}, |y|)}\notag\\
    &~~~~\times  e^{ - \frac{y^2}{4}\int^{\infty}_{X^+} d X_1^{+}\hat{q}_{k/A_2}(X_1^+, \mathbf{X}((X_1^+-X^+)/n_3^+ )- \mathbf{b}, |y|) - \frac{y^2}{4}\int^{\infty}_{X^-} d X_1^{-}\hat{q}_{k/A_1}(X_1^-, \mathbf{X}((X_1^--X^-)/n_3^-), |y|)}.
\end{align}
It factorizes into a convolution of a hard cross section and a medium-modified initial distributions, which is a mixture of the two incoming partons, as well as a medium-modified jet function. More in detail:
\begin{itemize}
    \item The initial line of eq.~(\ref{eq:kV_sum}) represents the probability of selecting a nucleon within a given nucleus, multiplied by a hard scattering factor that quantifies the interaction between parton $i$ and $j$ from those nucleons. The distributions of nucleons in nuclei $A_1$ and $A_2$ are represented by the functions $\rho_{A_{1/2}}$.
    
    \item The initial-state and final-state distributions are modified by the introduction of Wilson lines in the last two lines of eq.~(\ref{eq:kV_sum}). Two of these Wilson lines are associated with the active partons of the hard process and are semi-infinite past pointing. These depend on the background field generated by the {\it other} nucleus. In the third line, we observe that the final state (future-pointing) Wilson lines depend on the background field generated by both nuclei. We note the relation between the Gaussian expectation value of correlators of Wilson lines and the form involving $\hat q$, as presented in ref.~\cite{Baier:1996sk}, where $\hat q_{k/A_n}$ is the jet quenching parameter of parton $k$ in the medium $A_n$ defined in eq.~(\ref{eq:JetQuenchingParameter}) and is related to the gluon distribution $xG(x,1/|\bf{x}|^2)$.    
\end{itemize}

\subsection{Discussion about factorization}

The following discussion will present the key points of this analysis.
First, in order to demonstrate the factorization into initial and final state distributions, we have shown that interference diagrams in which one background gluon connects to an initial state parton in the amplitude and one background gluon connects to a final state parton in the conjugate amplitude are power-suppressed, alike in~\cite{Collins:1983ju}. This relies on two key features: 
(a) The hard process in coordinate space is localized to be at the same coordinate in both the amplitude and the conjugate amplitude. This is discussed in more detail in the section surrounding eq.~(\ref{eq:sigma0Forsingle}).
(b) The interaction with the background is mediated by a Glauber gluon, which occurs instantaneously. This feature of the Glauber gluon can be observed in eq.~(\ref{eq:DFGlauber}). Consequently, the possibility of interference diagrams is eliminated by (a) and (b), and the corresponding contributions to the observable are suppressed. It is anticipated that corrections to this result will be suppressed by large ratios of the following strongly ordered scales: $\Lambda_{QCD} \ll Q_T \ll Q$.
The observable can be described as an incoherent sum of initial and final state interactions.

Second, a review of the interaction between the active and spectator partons is now in order.
In both the Drell-Yan process and the photon-jet process, the initial or final distribution involves Wilson lines, which are process dependent. This process dependence implies the breakdown of collinear factorization, a concept often discussed in the TMD literature. In this case, such factorization is labeled as generalized factorization (see, for example, the references~\cite{Collins:2007nk,Rogers:2010dm,Mulders:2011zt}).
For a given parton from a nucleon, the Wilson line encodes the scattering with a background field generated by the partons belonging to the other nucleus.
This conclusion is also reached for the Wilson lines in the final state, where the background is generated by partons from both nuclei. This is in contrast to the case of Wilson lines that sum collinear gluons belonging to the same nucleus.

Finally, in order to achieve a comprehensive understanding, we must consider the influence of spectator-spectator interactions as it is done in the context of collinear factorization.
In collinear factorization, a significant amount of effort is dedicated to demonstrating how loop momenta involved in spectator-spectator topologies can be deformed out of the Glauber region. 
For a plus moving nucleon, there is a pinch singularity which originates from the spectator propagator pole and the active parton propagator pole in the minus momentum complex plane. In order to accommodate the deformation of the integration contour, it is necessary to remove the spectator's pole. This can be accomplished subsequent to the summation of all potential final-state cuts. Following this removal, the integration contour can be deformed out of the Glauber region.
For large values of transverse momentum flowing from the spectator to the active parton, and thus passing through the nucleon-active-spectator vertex, the wave function of the nucleon has the effect of suppressing the contribution. For further discussion, we direct the reader to the work~\cite{Collins:1982wa} and references therein.
In the present case, where we focus on the spectators coming from different nucleons which are those enhanced by the number of nucleons $A$, the spectator-spectator interactions do not involve a loop momentum since we assume the nucleons to be uncorrelated. Nevertheless, the same line of reasoning can be applied. In particular, in order for a spectator-spectator interaction to contribute to a significant momentum imbalance, $Q_T \gg \Lambda_{QCD}$, this requires a large momentum to flow through the nucleon-active-spectator vertex. This contribution is once again suppressed by the nucleon wave function, and thus is not accounted for in this analysis.

\section{Summary and Outlook}
\label{sec:conclusions}

In this work we have addressed the factorization of hard processes in nuclear collisions, directly assumed in most studies devoted to this subject, and its modification due to the interaction with other nucleons than those entering the hard scattering, called cold nuclear matter effects. We have focused on two observables: Drell-Yan and $\gamma$-jet. The nuclei have been modelled, following the Glauber model, as a superposition of independent nucleons, neglecting any interaction among them.

We have worked at the lowest order in perturbation theory, considering just one gluon exchange between the active and spectator partons and not including radiation. Then we have extended the study to multiple gluon exchanges. We have restricted to the kinematic situation where the final total transverse momentum of the observed Drell-Yan or $\gamma$-jet pair is much larger than the imbalance, in turn much larger than any non-perturbative scale. We have discarded power corrections of these scales, and considered only those contributions enhanced by the number of nucleons.

We have arrived at a form of factorization for the cross sections where the parton densities of the projectile and target become dressed by Wilson lines that depend on the other nucleons in both nuclei, thus being process dependent. The fields entering the Wilson lines can be described by the transport coefficient of nuclear matter used in many studies of the field, leading to broadening. On the other hand, factorization between initial state and final state effects is preserved at leading power. This kind of factorization is alike the generalized factorization discussed in the TMD literature.

The obvious outlook of our work is the inclusion of radiation from active initial and final partons and the link with the respective evolution equations and with the phenomenon of radiative energy loss leading to jet quenching. This further step, of considerable technical difficulty, is left for the future. As an example, in the present work the field in the Wilson lines that contain the cold nuclear matter effects, identical to those in DIS and pA~\cite{Kovchegov:1998bi, Kovchegov:2013cva,Kovchegov:2015zha}, results from the exchange of Glauber gluons with nucleons of both nuclei. It does not include contributions from two Glauber gluons exchanged from two nucleons from the same or different nuclei that after merging join a line entering the hard scattering (see, e.g.,~\cite{Kovchegov:2005ss, Wu:2017rry} for the resulting background gluon field). That is, in the background field that couples to the collinear partons participating in the hard collision (cf. eq.~(\ref{eq:FeynmanRules})):
\begin{equation}
A^\mu = A^\mu_{1,cl} + A^\mu_{2,cl} + A^\mu_{11/22/12,CGC} + \cdots,
\end{equation}
we restrict ourselves to the first two terms on the r.h.s, neglecting corrections involving the collision of one pair of nucleons, denoted as $A^\mu_{11/22/12,CGC}$.

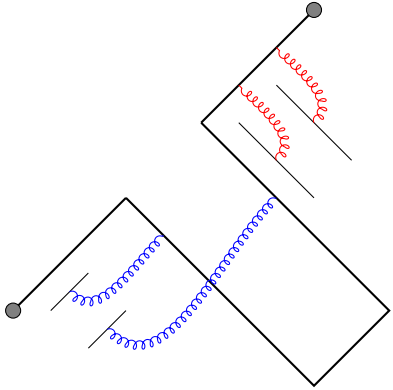
\begin{figure}
\begin{center}
    \begin{tikzpicture}[scale=0.5]
    \draw[thick] (-1,-1) -- (-4,-4);
    \draw[fill=black!50!white] (-4,-4) circle (0.2);
    \draw[thick] (1,1) -- (4,4);
    \draw[fill=black!50!white] (4,4) circle (0.2);
    \draw[thick] (-1,-1) -- ++(5,-5) -- ++(2,2) -- ++(-5,5);
    \draw (-3,-4)  -- ++(1,1);
    \draw[gluon,blue] (-2.5,-3.5) to[out=-45,in=-135] (0,-2);
    \draw (-2,-5)  -- ++(1,1);
    \draw[gluon,blue] (-1.5,-4.5) to[out=-45,in=-135] (3,-1);
    \draw (4,-1) -- ++(-2,2);
    \draw[gluon,red] (3,0) to[out=45,in=-45] (2,2);
    \draw (5,0) -- ++(-2,2);
    \draw[gluon,red] (4,1) to[out=45,in=-45] (3,3);
\end{tikzpicture} 
\end{center}
\caption{ Sketch of Wilson lines to be included in an operator definition of the PDF (in the direction $n_1$) containing cold nuclear matter effects. Gluons in blue are collinear to the $n_1$ direction, and gluons in red are Glauber gluons produced by nucleons moving in the $n_2$ direction. The two bullets indicate the location of the quark field insertions.}
\label{fig:operatordefinition}
\end{figure}

One additional point to be clarified is the operation definition of the new objects, modified TMD PDFs and jet functions, that we find. A tentative idea would be the inclusion of additional Wilson lines. For example, for DY, on top of the standard Wilson lines resumming collinear exchanges point to $-\infty$ in the direction of one nucleus, there would be Wilson lines resumming Glauber exchanges in the direction of the other nucleus, with longitudinal extent proportional to its length, see fig.~\ref{fig:operatordefinition}. The corresponding average should then be done on the wave functions of both nuclei.

\section*{Acknowledgments}

We thank Jianwei Qiu, Carlos Salgado, Matt Sievert and, specially, Yuri Kovchegov, for most useful comments. This work is part of the project CEX2023-001318-M financed by MCIN/AEI/\-10.13039/\-501100011033. We also acknowledge financial support from Xunta de Galicia (Centro singular de investigaci\'on
de Galicia accreditation 2019-2022, ref. ED421G-2019/05), by European Union ERDF, and by the Spanish
Research State Agency under project PID2020-119632GBI00.
B.W. acknowledges the support of the Ram\'{o}n y Cajal program with the Grant No. RYC2021-032271-I and the support of Xunta de Galicia under the ED431F 2023/10 project.
This work has been performed
in the framework of the European Research Council project ERC-2018-ADG-835105 YoctoLHC
and the MSCA RISE 823947 "Heavy ion collisions: collectivity and precision in saturation
physics" (HIEIC), and has received funding from the European Union’s Horizon 2020 research
and innovation programme under grant agreement No. 824093.

\bibliographystyle{JHEP}
\bibliography{jets.bib}

\end{document}